\documentclass[%
 reprint,
 amsmath,amssymb,
 aps,
nofootinbib
]{revtex4-1}

\usepackage{multirow}
\usepackage{amsmath}
\usepackage{booktabs}
\usepackage{tasks}
\usepackage{siunitx}
\usepackage{graphicx}
\usepackage{tabularx}% Include figure files
\usepackage{dcolumn}% Align table columns on decimal point
\usepackage{bm}% bold math
\usepackage[mathlines]{lineno}% Enable numbering of text and display math
\AtBeginDocument{%
  \heavyrulewidth=.08em
  \lightrulewidth=.05em
  \cmidrulewidth=.03em
  \belowrulesep=.65ex
  \belowbottomsep=0pt
  \aboverulesep=.4ex
  \abovetopsep=0pt
  \cmidrulesep=\doublerulesep
  \cmidrulekern=.5em
  \defaultaddspace=.5em
}

\usepackage{xcolor}

\graphicspath{{./}{Figures/}}

\begin{document}

%%%%%%%%%%%%%%%%%%%%%%%%%%%%%%%%%%%%%%%%%%%%%%%%%%%%%%%%%%%
\title{Search for neutrino lines from dark matter annihilation and decay with IceCube}

%%%%%%%%%%%%%%%%%%%%%%%%%%%%%%%%%%%%%%%%%%%%%%%%%%%%%%%%%%%
\affiliation{III. Physikalisches Institut, RWTH Aachen University, D-52056 Aachen, Germany}
\affiliation{Department of Physics, University of Adelaide, Adelaide, 5005, Australia}
\affiliation{Dept. of Physics and Astronomy, University of Alaska Anchorage, 3211 Providence Dr., Anchorage, AK 99508, USA}
\affiliation{Dept. of Physics, University of Texas at Arlington, 502 Yates St., Science Hall Rm 108, Box 19059, Arlington, TX 76019, USA}
\affiliation{CTSPS, Clark-Atlanta University, Atlanta, GA 30314, USA}
\affiliation{School of Physics and Center for Relativistic Astrophysics, Georgia Institute of Technology, Atlanta, GA 30332, USA}
\affiliation{Dept. of Physics, Southern University, Baton Rouge, LA 70813, USA}
\affiliation{Dept. of Physics, University of California, Berkeley, CA 94720, USA}
\affiliation{Lawrence Berkeley National Laboratory, Berkeley, CA 94720, USA}
\affiliation{Institut f{\"u}r Physik, Humboldt-Universit{\"a}t zu Berlin, D-12489 Berlin, Germany}
\affiliation{Fakult{\"a}t f{\"u}r Physik {\&} Astronomie, Ruhr-Universit{\"a}t Bochum, D-44780 Bochum, Germany}
\affiliation{Universit{\'e} Libre de Bruxelles, Science Faculty CP230, B-1050 Brussels, Belgium}
\affiliation{Vrije Universiteit Brussel (VUB), Dienst ELEM, B-1050 Brussels, Belgium}
\affiliation{Department of Physics and Laboratory for Particle Physics and Cosmology, Harvard University, Cambridge, MA 02138, USA}
\affiliation{Dept. of Physics, Massachusetts Institute of Technology, Cambridge, MA 02139, USA}
\affiliation{Dept. of Physics and The International Center for Hadron Astrophysics, Chiba University, Chiba 263-8522, Japan}
\affiliation{Department of Physics, Loyola University Chicago, Chicago, IL 60660, USA}
\affiliation{Dept. of Physics and Astronomy, University of Canterbury, Private Bag 4800, Christchurch, New Zealand}
\affiliation{Dept. of Physics, University of Maryland, College Park, MD 20742, USA}
\affiliation{Dept. of Astronomy, Ohio State University, Columbus, OH 43210, USA}
\affiliation{Dept. of Physics and Center for Cosmology and Astro-Particle Physics, Ohio State University, Columbus, OH 43210, USA}
\affiliation{Niels Bohr Institute, University of Copenhagen, DK-2100 Copenhagen, Denmark}
\affiliation{Dept. of Physics, TU Dortmund University, D-44221 Dortmund, Germany}
\affiliation{Dept. of Physics and Astronomy, Michigan State University, East Lansing, MI 48824, USA}
\affiliation{Dept. of Physics, University of Alberta, Edmonton, Alberta, Canada T6G 2E1}
\affiliation{Erlangen Centre for Astroparticle Physics, Friedrich-Alexander-Universit{\"a}t Erlangen-N{\"u}rnberg, D-91058 Erlangen, Germany}
\affiliation{Physik-department, Technische Universit{\"a}t M{\"u}nchen, D-85748 Garching, Germany}
\affiliation{D{\'e}partement de physique nucl{\'e}aire et corpusculaire, Universit{\'e} de Gen{\`e}ve, CH-1211 Gen{\`e}ve, Switzerland}
\affiliation{Dept. of Physics and Astronomy, University of Gent, B-9000 Gent, Belgium}
\affiliation{Dept. of Physics and Astronomy, University of California, Irvine, CA 92697, USA}
\affiliation{Karlsruhe Institute of Technology, Institute for Astroparticle Physics, D-76021 Karlsruhe, Germany }
\affiliation{Karlsruhe Institute of Technology, Institute of Experimental Particle Physics, D-76021 Karlsruhe, Germany }
\affiliation{Dept. of Physics, Engineering Physics, and Astronomy, Queen's University, Kingston, ON K7L 3N6, Canada}
\affiliation{Department of Physics {\&} Astronomy, University of Nevada, Las Vegas, NV, 89154, USA}
\affiliation{Nevada Center for Astrophysics, University of Nevada, Las Vegas, NV 89154, USA}
\affiliation{Dept. of Physics and Astronomy, University of Kansas, Lawrence, KS 66045, USA}
\affiliation{Department of Physics and Astronomy, UCLA, Los Angeles, CA 90095, USA}
\affiliation{Centre for Cosmology, Particle Physics and Phenomenology - CP3, Universit{\'e} catholique de Louvain, Louvain-la-Neuve, Belgium}
\affiliation{Department of Physics, Mercer University, Macon, GA 31207-0001, USA}
\affiliation{Dept. of Astronomy, University of Wisconsin{\textendash}Madison, Madison, WI 53706, USA}
\affiliation{Dept. of Physics and Wisconsin IceCube Particle Astrophysics Center, University of Wisconsin{\textendash}Madison, Madison, WI 53706, USA}
\affiliation{Institute of Physics, University of Mainz, Staudinger Weg 7, D-55099 Mainz, Germany}
\affiliation{Department of Physics, Marquette University, Milwaukee, WI, 53201, USA}
\affiliation{Institut f{\"u}r Kernphysik, Westf{\"a}lische Wilhelms-Universit{\"a}t M{\"u}nster, D-48149 M{\"u}nster, Germany}
\affiliation{Bartol Research Institute and Dept. of Physics and Astronomy, University of Delaware, Newark, DE 19716, USA}
\affiliation{Dept. of Physics, Yale University, New Haven, CT 06520, USA}
\affiliation{Columbia Astrophysics and Nevis Laboratories, Columbia University, New York, NY 10027, USA}
\affiliation{Dept. of Physics, University of Oxford, Parks Road, Oxford OX1 3PU, UK}
\affiliation{Dipartimento di Fisica e Astronomia Galileo Galilei, Universit{\`a} Degli Studi di Padova, 35122 Padova PD, Italy}
\affiliation{Dept. of Physics, Drexel University, 3141 Chestnut Street, Philadelphia, PA 19104, USA}
\affiliation{Physics Department, South Dakota School of Mines and Technology, Rapid City, SD 57701, USA}
\affiliation{Dept. of Physics, University of Wisconsin, River Falls, WI 54022, USA}
\affiliation{Dept. of Physics and Astronomy, University of Rochester, Rochester, NY 14627, USA}
\affiliation{Department of Physics and Astronomy, University of Utah, Salt Lake City, UT 84112, USA}
\affiliation{Oskar Klein Centre and Dept. of Physics, Stockholm University, SE-10691 Stockholm, Sweden}
\affiliation{Dept. of Physics and Astronomy, Stony Brook University, Stony Brook, NY 11794-3800, USA}
\affiliation{Dept. of Physics, Sungkyunkwan University, Suwon 16419, Korea}
\affiliation{Institute of Physics, Academia Sinica, Taipei, 11529, Taiwan}
\affiliation{Dept. of Physics and Astronomy, University of Alabama, Tuscaloosa, AL 35487, USA}
\affiliation{Dept. of Astronomy and Astrophysics, Pennsylvania State University, University Park, PA 16802, USA}
\affiliation{Dept. of Physics, Pennsylvania State University, University Park, PA 16802, USA}
\affiliation{Dept. of Physics and Astronomy, Uppsala University, Box 516, S-75120 Uppsala, Sweden}
\affiliation{Dept. of Physics, University of Wuppertal, D-42119 Wuppertal, Germany}
\affiliation{Deutsches Elektronen-Synchrotron DESY, Platanenallee 6, 15738 Zeuthen, Germany }

\author{R. Abbasi}
\affiliation{Department of Physics, Loyola University Chicago, Chicago, IL 60660, USA}
\author{M. Ackermann}
\affiliation{Deutsches Elektronen-Synchrotron DESY, Platanenallee 6, 15738 Zeuthen, Germany }
\author{J. Adams}
\affiliation{Dept. of Physics and Astronomy, University of Canterbury, Private Bag 4800, Christchurch, New Zealand}
\author{S. K. Agarwalla}
\thanks{also at Institute of Physics, Sachivalaya Marg, Sainik School Post, Bhubaneswar 751005, India}
\affiliation{Dept. of Physics and Wisconsin IceCube Particle Astrophysics Center, University of Wisconsin{\textendash}Madison, Madison, WI 53706, USA}
\author{J. A. Aguilar}
\affiliation{Universit{\'e} Libre de Bruxelles, Science Faculty CP230, B-1050 Brussels, Belgium}
\author{M. Ahlers}
\affiliation{Niels Bohr Institute, University of Copenhagen, DK-2100 Copenhagen, Denmark}
\author{J.M. Alameddine}
\affiliation{Dept. of Physics, TU Dortmund University, D-44221 Dortmund, Germany}
\author{N. M. Amin}
\affiliation{Bartol Research Institute and Dept. of Physics and Astronomy, University of Delaware, Newark, DE 19716, USA}
\author{K. Andeen}
\affiliation{Department of Physics, Marquette University, Milwaukee, WI, 53201, USA}
\author{G. Anton}
\affiliation{Erlangen Centre for Astroparticle Physics, Friedrich-Alexander-Universit{\"a}t Erlangen-N{\"u}rnberg, D-91058 Erlangen, Germany}
% ...

%%%%%%%%%%%%%%%%%%%%%%%%%%%%%%%%%%%%%%%%%%%%%%%%%%%%%%%%%%%
\begin{abstract}
Dark Matter particles in the Galactic Center and halo can annihilate or decay into a pair of neutrinos producing a monochromatic flux of neutrinos. The spectral feature of this signal is unique and it is not expected from any astrophysical production mechanism. Its observation would constitute a dark matter smoking gun signal. We performed the first dedicated search with a neutrino telescope for such signal, by looking at both the angular and energy information of the neutrino events.
To this end, a total of five years of IceCube's DeepCore data has been used to test dark matter masses ranging from 10~GeV to 40~TeV. No significant neutrino excess was found and upper limits on the annihilation cross section, as well as lower limits on the dark matter lifetime, were set. The limits reached are of the order of $10^{-24}$~cm$^3/s$ for an annihilation and up to $10^{27}$ seconds for decaying Dark Matter.
Using the same data sample we also derive limits for dark matter annihilation or decay into a pair of Standard Model charged particles.
\end{abstract}


\newpage

%%%%%%%%%%%%%%%%%%%%%%%%%%%%%%%%%%%%%%%%%%%%%%%%%%%%%%%%%%%
\maketitle
%%%%%%%%%%%%%%%%%%%%%%%%%%%%%%%%%%%%%%%%%%%%%%%%%%%%%%%%%%%

%%%%%%%%%%%%%%%%%%%%%%%%%%%%%%%%%%%%%%%%%%%%%%%%%%%%%%%%%%%
\section{Introduction}\label{sec:Introduction}
The existence of Dark Matter (DM) in the Milky Way and beyond can be probed indirectly through the observation of various kinds of particle fluxes produced by its annihilation or decay~\cite{review1, review2, review3}.
Looking at this possibility with neutrinos is of special interest because, as opposed to charged cosmic rays, neutrinos can propagate over very long distances without being deflected by magnetic fields. 
Moreover, neutrinos are much less absorbed than photons, in particular by the interstellar medium at very high energies or in dense environments.
Because of these two properties neutrinos point back to their origin of emission, even if this origin is far away or very dense, such as the center of the Sun or the center of the Earth~\cite{IceCube_Sun:2017,ANTARES_Sun:2016,IceCube_Earth}.

In this article, we search for monochromatic fluxes of neutrinos that could have been emitted by annihilation or decay of DM in the Milky Way. Monochromatic fluxes of neutrinos are produced when the parent particle is non-relativistic (as expected in the inner part of the Galaxy) and those processes proceeds into two-body final states where at least one of the two particles is a neutrino.
In the following we will assume that the second particle is also a neutrino, i.e.,~we consider the $\chi\chi \rightarrow \nu \overline{\nu}$ and $\chi \rightarrow \nu \overline{\nu}$ processes. 
The results we will get under this assumption can also be used for the case where the final state would consist of a single neutrino and another unknown particle, modulo a factor 2 for the neutrino flux and taking into account the shift of the neutrino line towards lower energies if the mass of this unknown particle is not negligible.
The neutrino line scenario is different from the usual setting where 
DM first creates a flux of Standard Model (SM) primary particles, leading to a continuum flux of secondary neutrinos from the subsequent decay (and hadronization) of these primary particles~\cite{IC86_GCWIMP_Event}.

The search of a monochromatic flux of neutrinos is interesting for at least four reasons.
First, when the signal has a narrow spectral feature it is easier to identify an excess against a broad continuous background stemming from atmospheric muons and neutrinos. This is of particular relevance for IceCube’s, so called, cascade events that comprises neutrinos with a better energy resolution compared to tracks~\cite{IceCube_reco}, (see Sec.~\ref{sec:Icecube} for details).  
Second, there is no high-energy astrophysical source that could mimic a monochromatic signal. Thus the observation of a line feature would constitute a DM {\it smoking gun} signal. 
Third, for the neutrino channel, neutrino telescopes can directly probe the primary neutrinos and thus have a clear advantage over gamma-ray telescopes (which in this case can only see secondary gamma-rays radiated by the neutrinos), despite the smaller neutrino interaction cross section. 
Finally, unlike $\gamma$-line production which generally proceeds at loop level (due to the electromagnetic neutrality of the DM particle, see e.g.~\cite{Garcia-Cely:2016hsk}), neutrino line production can proceed at tree level.
Systematic lists of simple tree level annihilation models can be found in~\cite{Lindner:2010rr,ElAisati:2017ppn}, and there exist numerous models where DM undergoes two body decays into neutrinos, see e.g.~\cite{Takayama:2000uz,Covi:2008jy,Hisano:2008ah,Guo:2010vy,Garny:2010eg,Feldstein:2013kka,Higaki:2014dwa,Rott:2014kfa,Esmaili:2014rma,Dudas:2014bca,ElAisati:2015qec,Garcia-Cely:2017oco,Patel:2019zky,Coy:2020wxp,Coy:2021sse,decay}, including models where the decay is induced by the neutrino mass seesaw interactions, see e.g.~\cite{Coy:2020wxp,Coy:2021sse,decay}.

IceCube has already performed searches of DM annihilating into a pair of neutrinos~\cite{IC86_GCWIMP}; however, these searches only used the angular information. 
In Ref.~\cite{ElAisati:2015ugc}, based on 2 years of public data the authors showed that given IceCube's energy resolution in cascade events, the search for a neutrino line with neutrino telescopes is clearly feasible and that the use of the energy information, crucial to distinguish a line from a continuum, allows for a clear improvement of the sensitivity. 
Here we use both the angular and energy distributions of the events, from five years of data consisting mostly of cascade events, with an optimized data selection for a monochromatic flux search from 10 GeV to 40 TeV.
An energy resolution for these cascade events of $\sim$~30\% for energies above 100 GeV is achieved.
Additionally, we also used the same data sample to get new constraints on an annihilation or decay into other charged particles. Again, the energy information of the events makes it possible to improve on the expected sensitivity. In total in this study we considered the annihilation and decay into the three neutrinos channels, $\nu_\mu \overline{\nu}_\mu$, $\nu_e \overline{\nu}_e$, $\nu_\tau \overline{\nu}_\tau$, and into a pair of $\tau^-\tau^+$,  $W^+W^-$, $\mu^+\mu^-$, and $b \overline{b}$ quarks. 

This article is structured as follows: Sec.~\ref{sec:Icecube} gives an overview of the IceCube Neutrino Observatory and the data selection used in this analysis. The signal expectation from dark matter annihilation and decay from the Galactic Center is described in Sec.~\ref{sec:signal}. In Sec.~\ref{sec:method} we describe the methodology and statistical tools used. Section~\ref{sec:sytematics} reviews the source of systematic uncertainties. Results are given in Sec.\ref{sec:results} for the neutrino channel and in Sec.~\ref{sec:results2} for other channels, and conclusions are given in Sec.~\ref{sec:conclusion}.

%%%%%%%%%%%%%%%%%%%%%%%%%%%%%%%%%%%%%%%%%%%%%%%%%%%%%%%%%%%
\section{IceCube and Data Selection}\label{sec:Icecube}

The IceCube Neutrino Observatory~\cite{IceCube} is a neutrino telescope located at the South Pole and buried between 1.5 and 2.5 km in the Antarctic ice sheet. It consists of a three-dimensional array of 5,160  photo-multipliers (PMTs) that detect the Cherenkov light induced by charge particles created in the neutrino interactions with the surrounding matter. The PMTs are housed in the Digital Optical Modules (DOMs), which also contain the electronics for the digitization of the signal~\cite{IceCube_DOM}. DOMs are separated 17~m vertically and 125~m horizontally to optimize the detection of TeV neutrinos. A denser sub-array of the IceCube detector with a reduced DOM spacing, called DeepCore~\cite{DeepCore}, is located at the bottom-center of the IceCube array and it is sensitive to neutrinos with energies ~$\gtrsim 10$~GeV. 
Depending on the neutrino interaction, different signatures can be observed within the instrumented volume of IceCube. Charge-current interactions of $\nu_\mu$ will leave a track-like signature, while neutral-current interactions of all flavors and charge-current interactions of $\nu_e$ will induce a hadronic or electromagnetic shower leaving a spherical light pattern in the detector, which we call cascades. The same is true for charge-current interactions of $\nu_\tau$, since at energies below 1 PeV the $\tau$ decay length is shorter than the average distance of IceCube's DOMs rendering the $\tau$ track undetectable.

The event selection used in this analysis was developed for DeepCore and was optimized to identify and select cascade events~\cite{IC86_GCWIMP_Event}. Filtered data uses DeepCore's fiducial volume while the neighboring strings of IceCube are used as a veto from atmospheric muons. The final selection uses Boosted Decision Trees (BDTs)~\cite{TMVA} trained with scrambled data as background and different reference signals from Monte Carlo (MC) simulations. Two different signal benchmarks were used: a neutrino spectra generated from a 100 GeV dark matter mass annihilating into $b\overline{b}$ and a 300 GeV mass dark matter particle annihilating into $W^+W^-$. Although in this analysis we focused on direct annihilation and decay into neutrinos, the choice of the two spectra used in BDTs was made to represent a soft and hard neutrino spectrum respectively. This improves the sensitivity over a wide range of masses as well as different spectra. 
The scores produced by the two BDTs were used to define two different event selections, one optimized for best sensitivity of the analysis for dark matter masses from 10 GeV to 1 TeV, and the other for masses from 1 TeV towards 40 TeV. 

The energy resolution, defined as the standard deviation of $E_{rec}/E_{true}$ distribution, improves from 60\% at 10 GeV to 30\% beyond 100 GeV. Both datasets also have a similar median angular resolution, ranging from $\sim 50^\circ$ for soft annihilation channels like $b\overline{b}$ to $\sim 20^\circ$ for the $W^+W^-$ annihilation channel at a dark matter mass of 300 GeV, which is sufficient to exploit the directional information for large extended regions of emission such as the Galactic dark matter halo.

%%%%%%%%%%%%%%%%%%%%%%%%%%%%%%%%%%%%%%%%%%%%%%%%%%%%%%%%%%%
\section{Signal Expectation}\label{sec:signal}

Since neutrinos can travel unhindered through the Galaxy, the neutrino energy spectrum remains almost identical to the spectrum at the production site. The differential neutrino flux from dark matter annihilation in an observational volume at angular distance towards the Galactic Center, $\Psi$, can be written as:

\begin{equation}
    \frac{\mathrm{d}\phi_{\mathrm{\nu}}}{\mathrm{d}E_{\mathrm{\nu}}} (E_\nu, \Psi) = \frac{1}{4\pi} \frac{\langle \sigma v \rangle}{2  m_{\chi}^2} \frac{\mathrm{d}N_{\mathrm{\nu}}}{\mathrm{d}E_{\mathrm{\nu}}} \int_{\Delta \Omega}  \int_{\mathrm{l.o.s.}} \rho_\chi^2\left(r(\ell, \Psi)\right) \mathrm{d}\ell \; \mathrm{d}\Omega ,
    \label{eq:annihilation_flux}
\end{equation}

\noindent where $\langle \sigma v \rangle$ is the thermally averaged dark matter self-annihilation cross-section, and $m_\chi$ is the dark matter mass. The factor 2 in the denominator assumes that DM is a Majorana particle and therefore its own anti-particle~\cite{DM_Majorana}.
Likewise, the differential neutrino flux from a decaying dark matter can be expressed in terms of the decaying lifetime, $\tau_\chi$, as: 

\begin{equation}
        \frac{\mathrm{d}\phi_{\mathrm{\nu}}}{\mathrm{d}E_{\mathrm{\nu}}} (E_\nu, \Psi) = \frac{1}{4\pi} \, \frac{1}{\tau_\chi m_\chi} \; \frac{\mathrm{d}N_{\mathrm{\nu}}}{\mathrm{d}E_{\mathrm{\nu}}} \int_{\Delta \Omega}  \int_{\mathrm{l.o.s.}} \rho_\chi\left(r(\ell, \Psi)\right)  \mathrm{d}\ell \;  \mathrm{d}\Omega.
    \label{eq:decay_flux}
\end{equation}

\begin{figure*}
 \centering
    \begin{minipage}{.49\textwidth}
    \includegraphics[width=\linewidth]{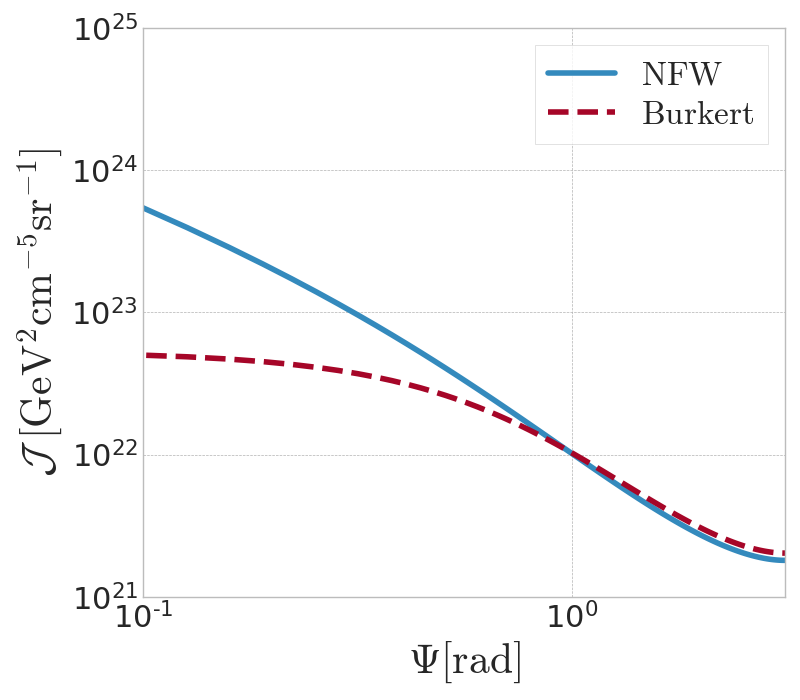}
    \end{minipage}
    \hfill
    \begin{minipage}{.49\textwidth}
    \includegraphics[width=\linewidth]{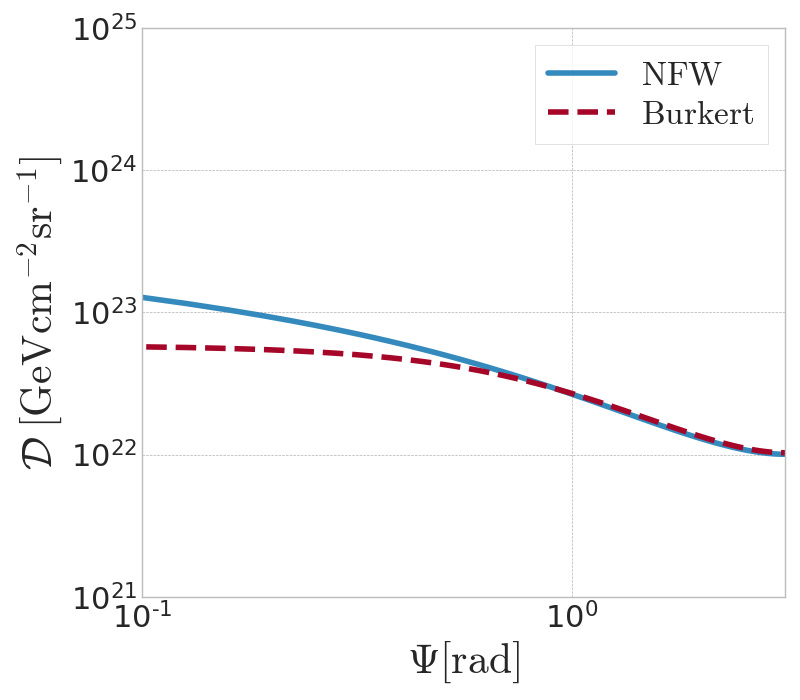}
    \end{minipage}
    
    \caption{Differential $J$-factor (left) and $D$-factor (right) as a function of the angular orientation towards the Galactic Center for two assumption of the dark matter density profile $\rho_\chi(r)$, the NFW~\cite{NFW} in solid blue and the Burkert~\cite{Burkert} profile in dotted red. }
    \label{fig:JFactors}
\end{figure*}

In both cases the last two integrals encompass all the astrophysical information given by the dark matter density distribution in the Milky Way, $\rho_\chi(r(\ell, \Psi))$, and they are usually referred as the $J$-factor and $D$-factor respectively (see Fig.~\ref{fig:JFactors}):

 \begin{equation}
 \begin{split}
       J{\rm -factor}& \equiv \int_{\Delta \Omega}  \mathcal{J} \; \mathrm{d}\Omega = \int_{\Delta \Omega}  \int_{\mathrm{l.o.s.}} \rho_\chi^2\left(r(\ell, \Psi)\right)  \mathrm{d}\ell  \;\mathrm{d}\Omega  ,\\
        D{\rm -factor}& \equiv  \int_{\Delta \Omega} \mathcal{D} \; \mathrm{d}\Omega  = \int_{\Delta \Omega}  \int_{\mathrm{l.o.s.}} \rho_\chi\left(r(\ell, \Psi)\right)  \mathrm{d}\ell \;\mathrm{d}\Omega .
\end{split}
\label{eq:j-d-factors}
 \end{equation}

The argument $r(\ell, \Psi)$ is the Galactocentric distance expressed as a function of the angle with respect to the Galactic Center $\Psi$ and integrated over the field-of-view $\Delta \Omega$, and $\ell$ which is the distance along the line of sight (l.o.s.).  
The dark matter density distribution is inferred from first principles, numerical simulations, and astronomical observation and it is subject to large uncertainties~\cite{Benito_2019}. As it is custom in indirect searches of dark matter, the signal predictions and results are evaluated for two different assumptions of the density profile. In this article, we used the Navarro-Frenk-White (NFW)~\cite{NFW} and Burkert~\cite{Burkert} profiles. Both of them assume a spherical dark matter distribution but with a different radial profile. As can be seen in Fig.~\ref{fig:JFactors}, the annihilation signal expectation is more impacted by the choice of the density profile than the decay signal, due to its dependency on $\rho^2_\chi$ in the $J$-factor compared to the $\rho_\chi$ dependency in the $D$-factor.

The term $\mathrm{d}N_{\mathrm{\nu}}/\mathrm{d}E_{\mathrm{\nu}}$ describes the number of neutrinos per energy unit, at a given energy $E_\nu$ produced per annihilation or decay at the source.
For the $\chi\chi \rightarrow \nu_\alpha\overline{\nu}_\alpha$ and $\chi \rightarrow \nu_\alpha\overline{\nu}_\alpha$ channels (with $\alpha = e, \mu, \tau$),
the spectra can be described by a $\delta$-function centered at $E_\nu = m_\chi$ and $E_\nu = m_\chi/2$ respectively. Note that electro-weak corrections increase the amount of low-energy neutrinos altering the mono-chromatic spectra in non-trivial ways. In this study, we used neutrino spectra calculated in PPPC4~\cite{Cirelli}, which includes electro-weak corrections at leading order as described in more details in~\cite{EW-wc} (for a more recent treatment of electro-weak corrections see~\cite{EW-wc2}). 
For the neutrino line channels the continuum of lower energy neutrinos induced by the electro-weak corrections\ is negligible for our analysis.
For an annihilation or decay into a pair of charged particles, the secondary neutrino spectra from parton showers and hadronization are also estimated using the tables provided in PPPC4~\cite{Cirelli}. 

For all channels studied a branching ratio of 100\% is assumed at the source, including those into each monochromatic neutrino flavor line. 
However, long baseline vacuum neutrino oscillations will produce similar amount of electron, muon, and tau neutrinos \cite{ElAisati:2015ugc,nu_scillations}.
Simplistically, we take all our neutrino signals to have a democratic flavor composition when arriving at the detector.\footnote{Since the monochromatic neutrino-line spectra are basically identical for all flavors, this is in practice the same as assuming that these signals have democratic flavor composition at the source.}

For the decay mode, as only one dark matter particle is needed to produce a decaying signal, an additional extra-Galactic and potential Galactic substructure components can have a sizable contribution. However, for angular distances of less than 30$^\circ$ with respect to the Galactic center, these extra components are typically negligible for both decaying and annihilating DM and are therefore not considered in this analysis~\cite{sub1, sub2}.

%%%%%%%%%%%%%%%%%%%%%%%%%%%%%%%%%%%%%%%%%%%%%%%%%%%%%%%%%%%
\section{Analysis Method}\label{sec:method}
In this analysis we used a Poisson binned likelihood method with two observables: the reconstructed energy of the event, $E_{rec}$, and the angular distance with respect to the Galactic Center, $\Psi_{rec}$. The likelihood expression can be written as:

\begin{equation}
\mathcal{L}(\mu) = \prod_{i=0}^{N_{bins}} {\rm Poisson}\left(n_i|N_{obs}^{total}\cdot f_i(\mu)\right) ,
\label{eg:likelihood}
\end{equation}

\noindent where maximization is performed against the signal fraction, $\mu = {n_s}/{N_{obs}^{total}} \in  [0,1]$, where $n_s$ is the number of signal events in the sample and $N_{obs}^{total}$ is the total number of events observed.  The latter is taken from data and therefore is not a free parameter of the model. For each bin in the energy-angular distance space, the expected number of events is given by $N_{obs}^{total}\cdot f_i(\mu)$, where $f_i(\mu)$ is the fraction of events falling in the $i$th bin, given by

\begin{equation}
    f_i(\mu) = (1-\mu) \cdot\mathcal{B}_i + \mu\cdot \mathcal{S}_i ,
\end{equation}

\noindent where $\mathcal{S}_i$ and $\mathcal{B}_i$ are the signal and background probability density functions ({\it pdf}), respectively. A common neutrino telescope procedure is to build the background model from experimental data by scrambling the right ascension coordinate. This technique consists on assigning a uniformly random distributed right ascension to the events in order to create a background pseudo-sample. This is possible since neutrino telescopes have a nearly constant duty cycle and as Earth's rotates the atmospheric neutrino and muon backgrounds become uniform in right ascension. Scrambling is a powerful technique but assumes that any potential signal is negligible and so it will be diluted by the scrambling of the right ascension. %This assumption is particularly valid for analyzing signal emitted by point-sources. However as the signal region occupies a large fraction of the sky, the scrambling technique becomes less efficient as the presence of a possible signal will affect the background estimate. 
In order to correct for a possible signal {\it contamination} in the background estimate, we make use of a signal subtraction likelihood~\cite{IC86_GCWIMP_Event}. In this case, the estimated background-only {\it pdf} can be written as

\begin{equation}
    \mathcal{B}_i(\mu) = \frac{1}{1-\mu}\left[ \mathcal{B}^{scrambled}_i - \mu \mathcal{S}^{scrambled}_i\right],
\label{eq:background_only}
\end{equation}

\noindent where $S^{scrambled}$ is the {\it pdf} of a right ascension scrambled signal computed from simulation. The final expression for the signal fraction can be written as

\begin{equation}
    f_i(\mu) = \mathcal{B}^{scrambled}_i + \mu\cdot \left(\mathcal{S}_i - \mathcal{S}^{scrambled}_i\right).
    \label{eq:fraction_signalsubtraction}
\end{equation}

\begin{figure*}
    \centering
    \begin{minipage}{.49\textwidth}
    \includegraphics[width=\linewidth]{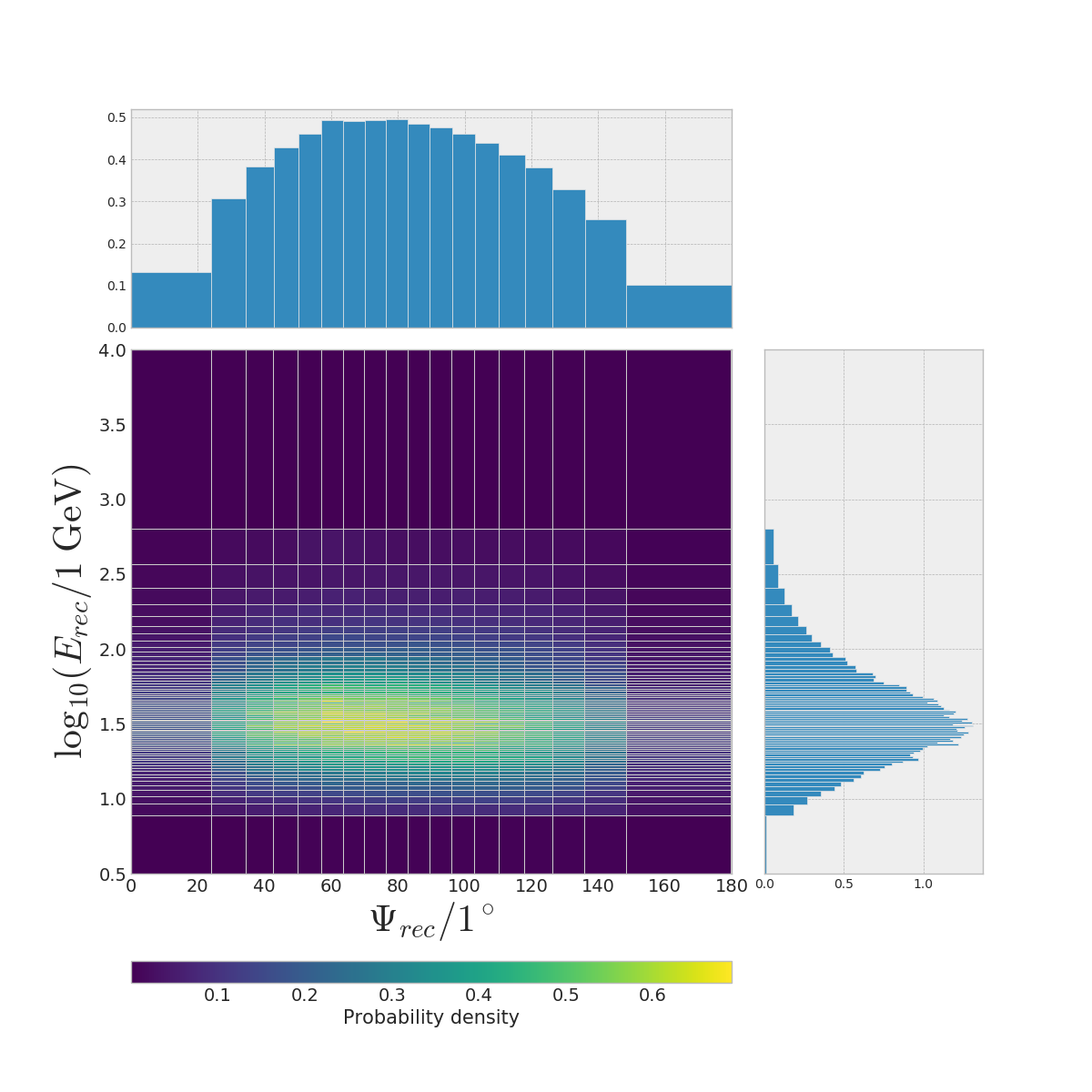}
    \end{minipage}
    \hfill
    \begin{minipage}{.49\textwidth}
    \includegraphics[width=\linewidth]{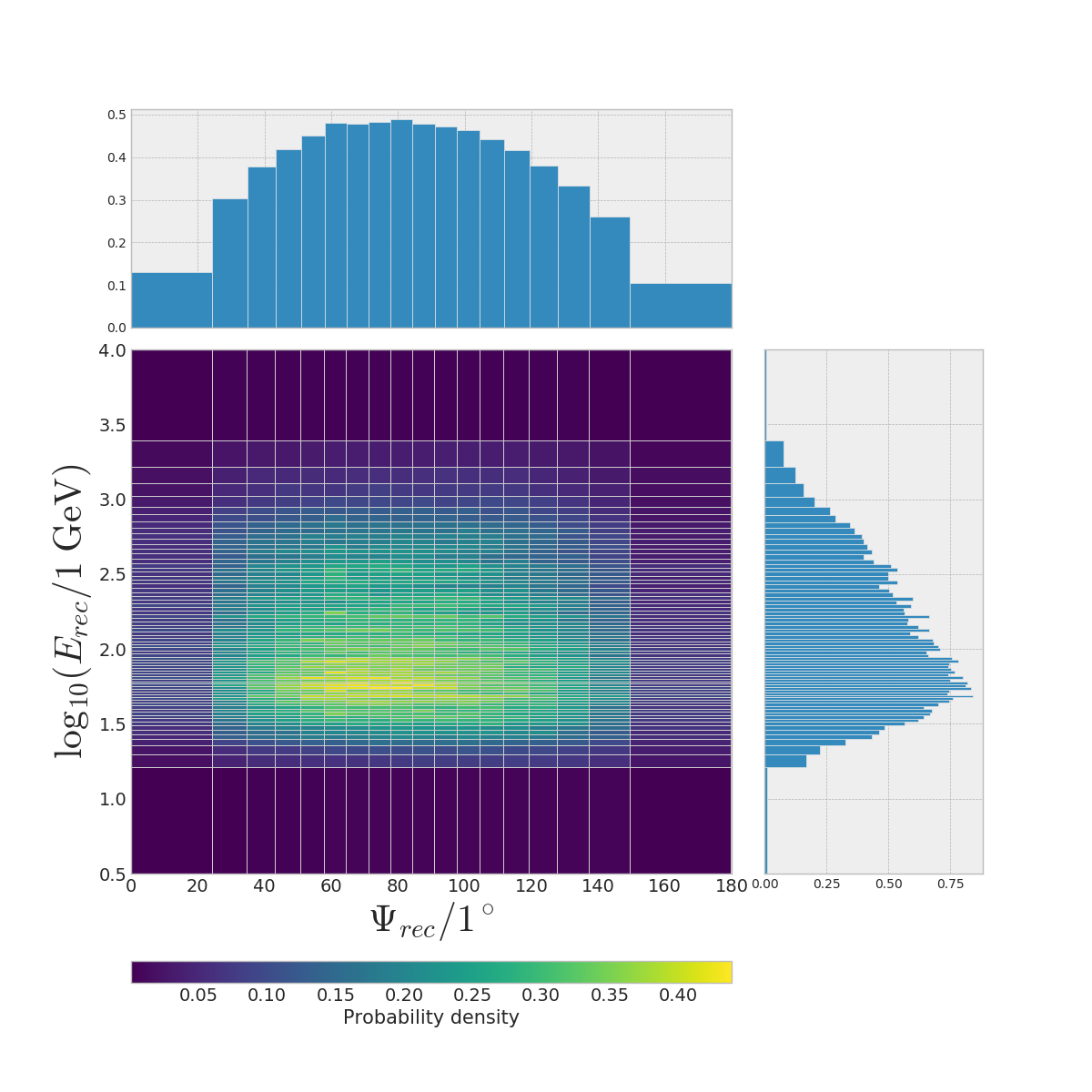}
    \end{minipage}
    \caption{Background probability density functions,
    for the low-energy (left) and high-energy selections (right) as functions of the reconstructed energy and the angular distance towards the Galactic Center. The non-uniform binning was used in order to ensure that no empty bins are present in the distributions.}
    \label{fig:Bkg_PDFs}
\end{figure*}

Fig.~\ref{fig:Bkg_PDFs} shows the background {\it pdf} built from an average of 100 right ascension scrambled pseudo samples for both the low-energy (left) and the high-energy selections (right). Data distributions have small number of events at the tails of the energy distributions. In order to avoid empty bins in the background {\it pdf}, which might be specially problematic for monochromatic signal expectations, we used a binning based on quantiles resulting in each bin containing roughly the same amount of events. This limits the statistical error per bin in the estimation of the {\it pdf} from scrambled data. The implementation of the quantile binning was done using the software {\tt physt}~\footnote{https://github.com/janpipek/physt/}. 

The binning used for the background {\it pdf} is then applied to the signal distributions. The signal {\it pdf's} are built by re-weighting neutrino MC simulations according to the expression given by equations~\ref{eq:annihilation_flux} and~\ref{eq:decay_flux}. These MC datasets include simulation of all three neutrino flavors. In order to reduce the impact of weighted MC errors, the simulation was oversampled by duplicating high weighted events at different arrival times. This technique produces a smooth signal distribution while preserving the energy and angular response of the detector. The signal depends on the dark matter mass, the halo profile, and the annihilation or decay channel. 

\begin{figure*}
    \centering
    \begin{minipage}{.49\textwidth}
    \includegraphics[width=\linewidth]{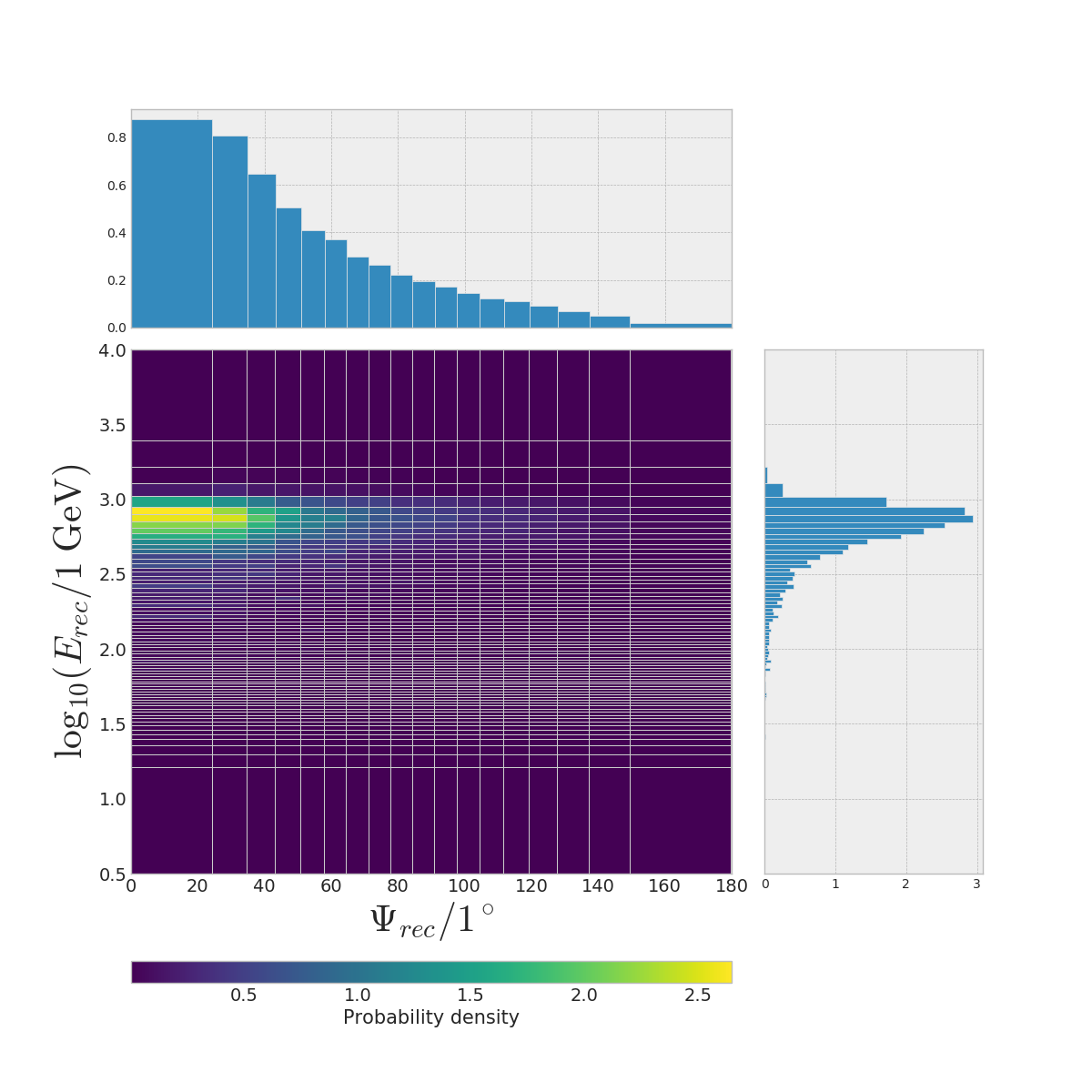}
    \end{minipage}
    \hfill
    \begin{minipage}{.49\textwidth}
    \includegraphics[width=\linewidth]{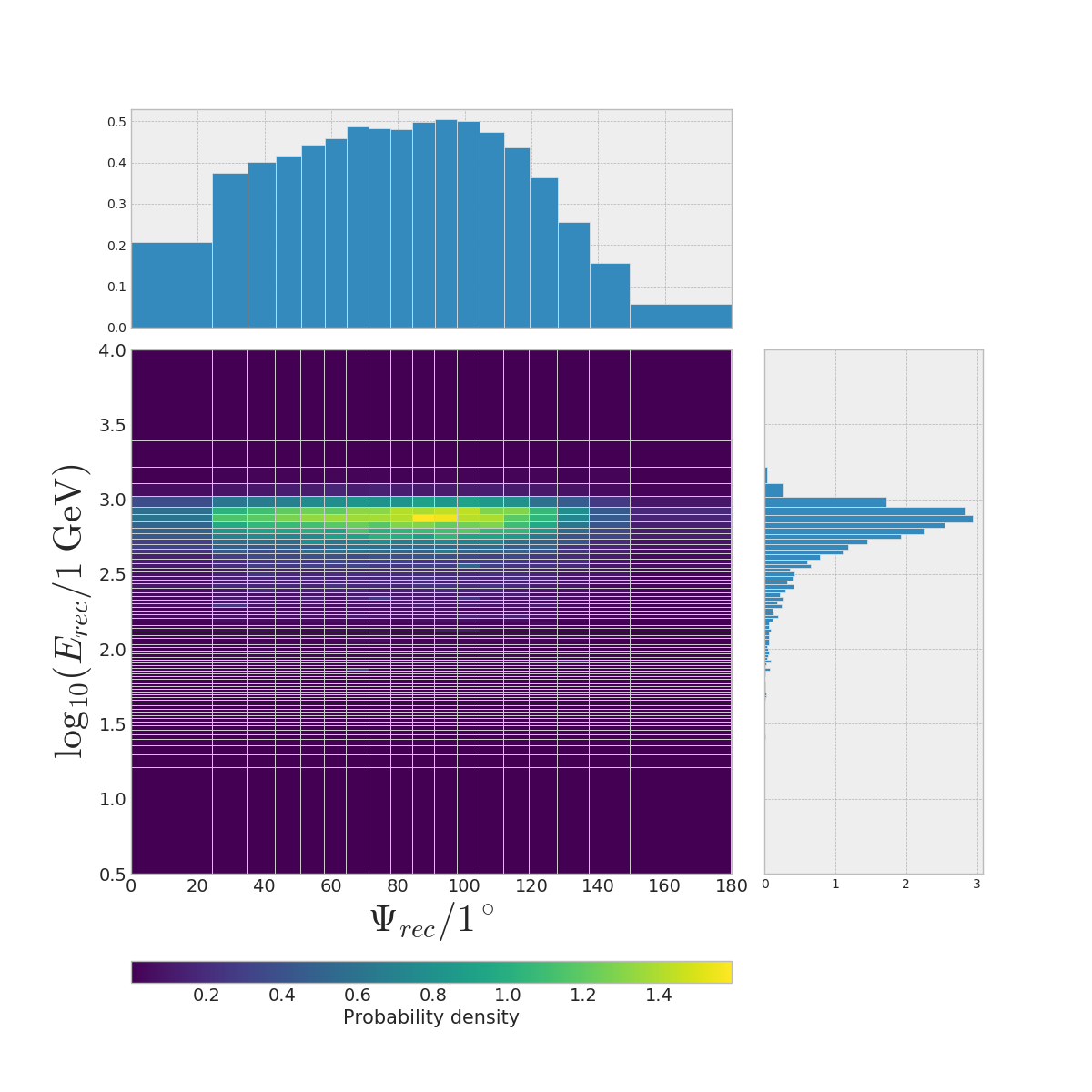}
    \end{minipage}
    \caption{Left: Signal probability density function 
    for a dark matter particle of $m_\chi = 1\; {\rm TeV}$ annihilating to the $\nu_e \overline{\nu}_e$ channel and the NFW profile for the high-energy selection. Right: same signal distribution but scrambled in right ascension.} 
    \label{fig:Signal_PDFs}
\end{figure*}

The left panel on Fig.~\ref{fig:Signal_PDFs} shows the signal {\it pdf} for the benchmark annihilation channel to $\nu_e \overline{\nu}_e$ and a dark matter mass of $1\; {\rm TeV}$. As can be seen in the projected distribution of reconstructed energy, the spectra features a sharp peak corresponding to the monochromatic signal. The right panel shows the same distribution but scrambled in right ascension. As expected from the scrambled method we consider, the projected distribution in reconstructed energy remains identical. The scrambled signal distribution is used in the minimization in order to correct the background {\it pdf} as shown in equation~\ref{eq:background_only}.

 In this analysis, we tested about 19 different values for the dark matter mass ranging from 10~GeV  to 40~TeV. The final number of masses was selected by verifying that an injected signal in-between two consecutive masses will be recovered, while at the same time limiting the computational hurdle of evaluating a large number of masses. As mentioned in Sec.~\ref{sec:Introduction}, in addition to the three neutrino channels, we also evaluate the neutrino signal coming from the $W^+W^-$, $\tau^+\tau^-$, $b\overline{b}$, and $\mu^+\mu^-$ channels for both annihilation and decaying dark matter and the two halo profiles. 
For each combination of mass, channel, and halo profile, the analysis finds the $\hat{\mu}$ that maximizes the likelihood. This best estimate can then be translated to a thermally averaged annihilation cross-section, $\langle \sigma v \rangle$, or a decaying dark matter lifetime $\tau_\chi$. The significance of the result, or compatibility with the null hypothesis $\mathcal{H}_0(\mu = 0)$, is calculated using the discovery test-statistics, $q_0$, defined as 
\begin{equation}
q_0 = \begin{cases} 
     -2\log\frac{\mathcal{L}(\hat{\mu})}{\mathcal{L}(\mu = 0)} & \hat{\mu}\geq 0 \\
      0 & \hat{\mu} < 0,  
      \end{cases}
\label{eq:ts_discovery}
\end{equation}
where we assume that the physical parameter must be positive, $\mu \ge 0$, so that null hypothesis can only be rejected when the data prefers a positive signal contribution.

%%%%%%%%%%%%%%%%%%%%%%%%%%%%%%%%%%%%%%%%%%%%%%%%%%%%%%%%%%%%%%%%%
\section{Systematic Uncertainties}\label{sec:sytematics}
Since the background {\it pdf} is built essentially from data, there are no systematic uncertainties affecting the shape of the background model. 
The influence on the scrambled signal {\it pdf} from systematic uncertainties only affects slightly the background estimate for very large signal fractions. 
Still, signals’ {\it pdf's} can have notable fluctuations, because they are based on a limited number of MC events and the technique of oversampling those.
By comparing the results from the practically identical signals $\nu_e \overline{\nu}_e$, $\nu_\mu \overline{\nu}_\mu$ and  $\nu_\tau \overline{\nu}_\tau$  ---recall that these neutrino-lines' energy spectra have negligibly different electro-weak corrections and democratic flavour compositions at the detector--- we see from the tables \ref{tb:nuenue}, \ref{tb:numunumu} and \ref{tb:nutaunutau} in the Appendix that limits are essentially unaffected, differing by few tens of percents and reaching a factor of three for the lowest mass.

Detector systematic uncertainties will, in addition, affect the efficiency of the detector and might introduce a bias in the fitted signal fraction, which will influence the conversion from estimated number of signal events or upper limits to the physical parameters, $\langle \sigma v \rangle$ and $\tau_\chi$. Among the known detector systematic uncertainties we evaluated the DOM efficiency and several ice properties. Variations in the detector parameters, within their systematic uncertainties, result in a 30\% uncertainty in the nominal detector sensitivity, below the statistical uncertainties due to fluctuations of the background. 
These effects on the result are, however, far much smaller than the effect due to astrophysical uncertainties. 
As usual for dark matter indirect searches the latter constitute the dominant source of uncertainty.
 
In particular, the shape of the dark matter halo profile can have a large impact on the results. For this reason we will consider two typical bracketing halo profiles. 

%%%%%%%%%%%%%%%%%%%%%%%%%%%%%%%%%%%%%%%%%%%%%%%%%%%%%%%%%%%
\section{Results for the neutrino line channel}\label{sec:results}

After performing the likelihood maximization on all the masses and neutrino channels for both halo profiles and for both the annihilation and the decay modes, no significant excess with respect to the background expectation is found. In the absence of such a signal we place upper limits on the thermally averaged annihilation cross-section and lower limits on the dark matter decay lifetime. In order to establish upper limits on the signal fraction we used the test-statistics defined as~\cite{Cowan2011-xc}:

\begin{equation}
q_\mu = \begin{cases} 
      -2\log\frac{\mathcal{L}(\hat{\mu})}{\mathcal{L}(\mu)} & \hat{\mu}\leq \mu \\
       0 & \hat{\mu} > \mu. 
      \end{cases}
\label{eq:ts_upperlimit}
\end{equation}

An upper limit is built by selecting the value $\mu$ producing a significance of 10\%, under the same $\mathcal{H}_1(\mu)$ hypothesis. After verifying that the asymptotic distribution of $q_\mu$ correctly follows a half $\chi^2$ distribution for one degree of freedom as dictated by a generalization \cite{Wilks2} of Wilks' theorem~\cite{Wilks}, we used the value of $q_\mu = 1.64$ to calculate the limits. 

\begin{figure}
    \centering
    \begin{minipage}{\linewidth}
    \includegraphics[width=\linewidth]{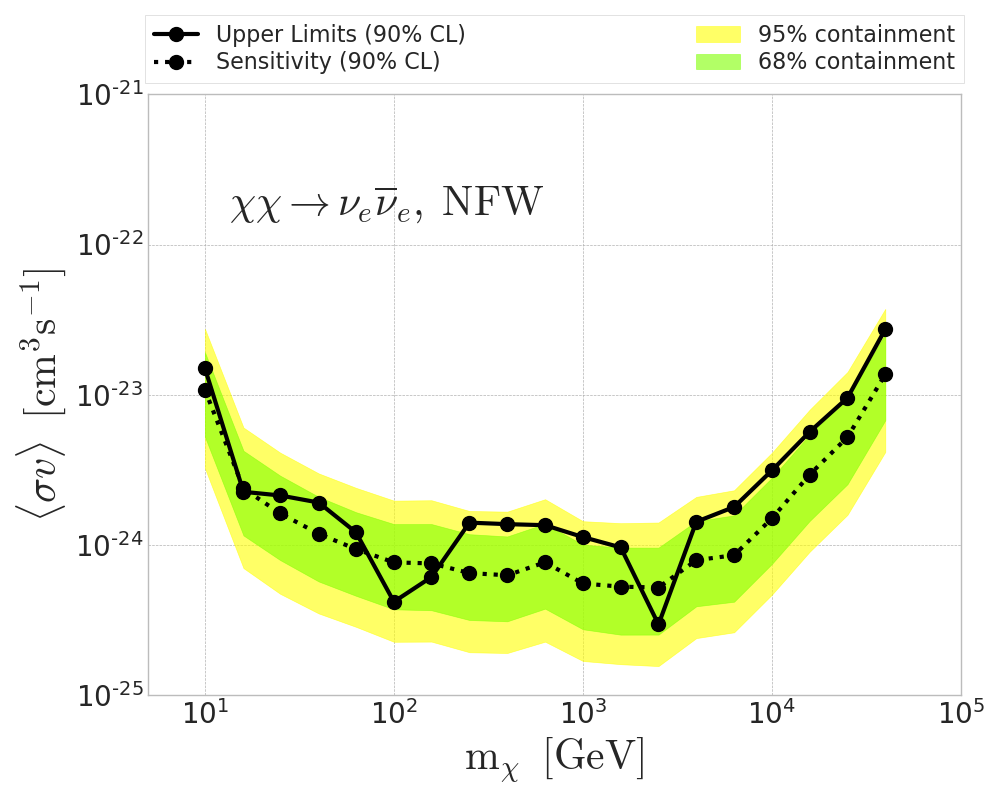}
    \end{minipage}
    \hfill
    \begin{minipage}{.49\textwidth}
    \includegraphics[width=\linewidth]{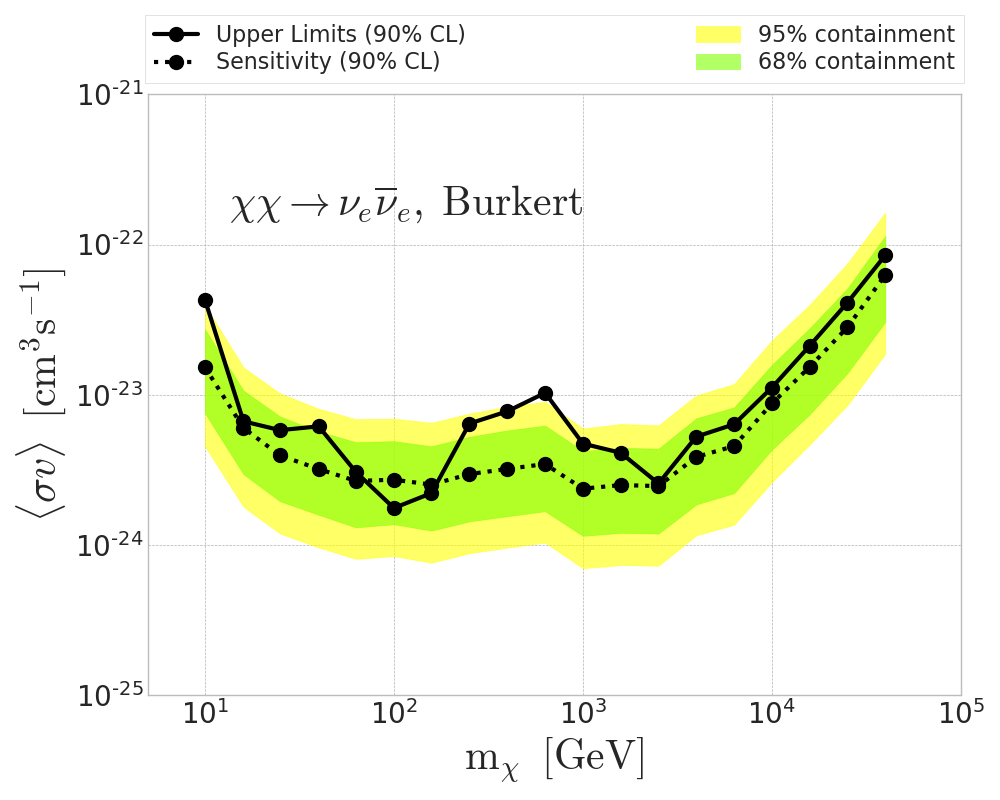}
    \end{minipage}
    \caption{Upper limits (solid line) and sensitivity (dotted line) at 90\% C.L. on the thermally averaged self-annihilation cross-section of the $\nu_e\overline{\nu}_e$ channel and NFW profile (top) and Burkert profile (bottom) as function of the dark matter mass together with the $1\sigma$ (green) and $2\sigma$ (yellow) containment bands for the expected sensitivity.} 
    \label{fig:results_annihilation_nuenue}
\end{figure}

Fig.~\ref{fig:results_annihilation_nuenue} shows the upper limits obtained on the thermally averaged annihilation cross-section for the $\nu_e \overline{\nu}_e$ final state assuming a NFW (top panel) or Burkert (bottom panel) dark matter halo profile as a function of the dark matter mass. The dotted line indicates the expected median upper limit, or sensitivity, in the absence of signal at one-sided 90\% C.L. while the green and yellow bands indicate the 68\% and 95\% expected background fluctuations. Upper limits are only evaluated at the corresponding mass points and lines in-between are only used to guide the eye. Masses below 1 TeV are evaluated with the low-energy dataset while larger masses are tested with the high-energy selection. There is a mild positive fluctuation towards $\sim 1$ TeV in dark matter masses for this neutrino channel. The local significance of this fluctuation does not exceed  $\sim$ 1.3\,$\sigma$ ($p$-value of $\sim$ 10\%) and it is visible in both profiles. Due to the quantile binning procedure, at higher energies there is a strong correlation among masses between 10 TeV to 200 TeV which explains why upper limits are consistently above the median sensitivity over such a broad range of masses. 

Similar results are obtained for the practically identical signals from the two other neutrino flavors, as shown in tables \ref{tb:nuenue} to \ref{tb:nutaunutau} in the Appendix.

\begin{figure}
    \centering
    \begin{minipage}{\linewidth}
    \includegraphics[width=\linewidth]{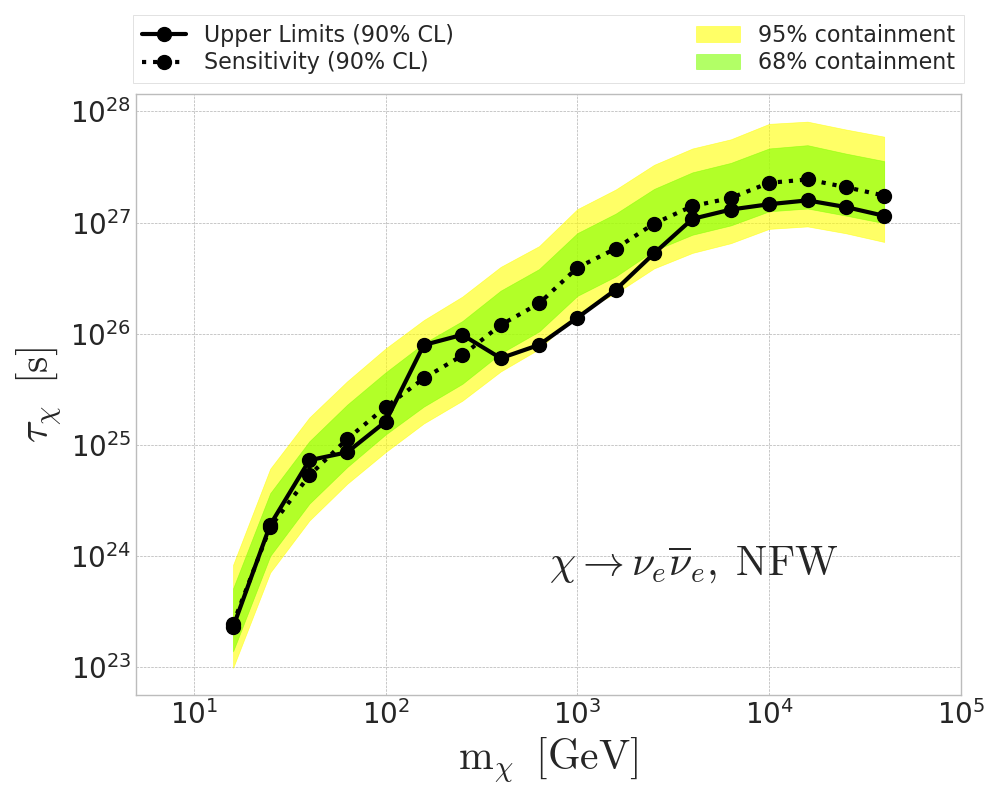}
    \end{minipage}
    \hfill
    \begin{minipage}{.49\textwidth}
    \includegraphics[width=\linewidth]{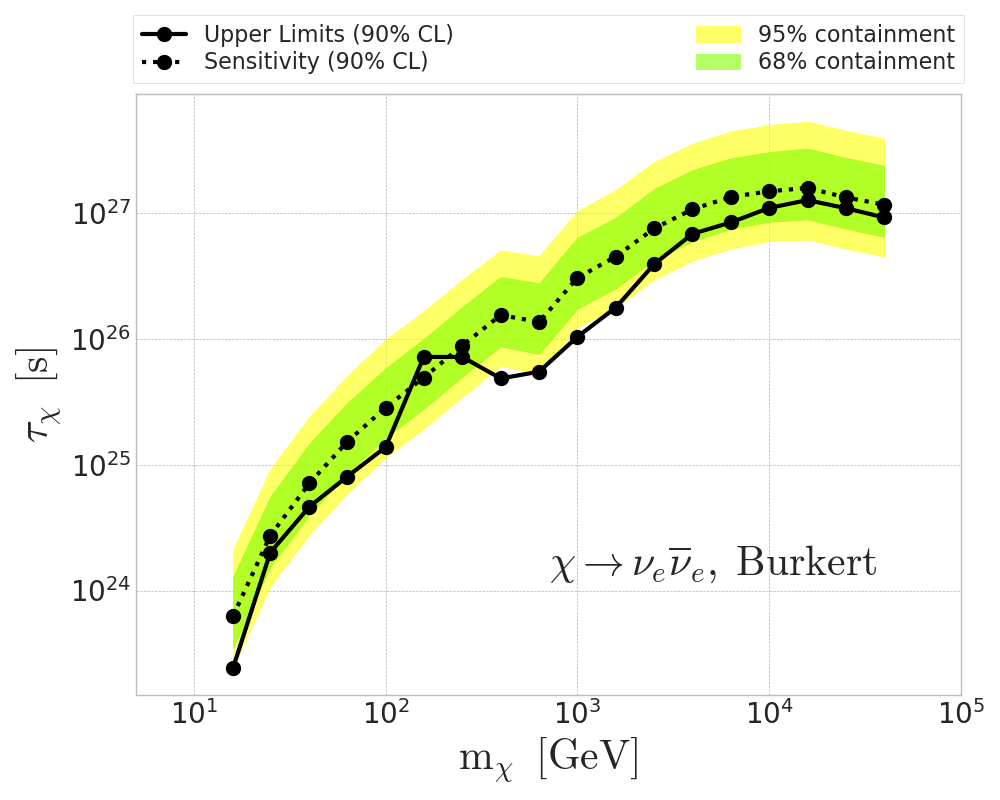}
    \end{minipage}
    \caption{Lower limits (solid line) and sensitivity (dotted line) at 90\% C.L. on the decaying dark matter lifetime $\tau_\chi$ of the $\nu_e\overline{\nu}_e$ channel for NFW profile (top) and Burkert profile (bottom) as function of the dark matter mass together with the $1\sigma$ (green) and $2\sigma$ (yellow) containment bands for the expected sensitivity.} 
    \label{fig:results_decay_nuenue}
\end{figure}

Results on DM decays for the same $\nu_e \overline{\nu}_e$ neutrino channel are summarized in Fig.~\ref{fig:results_decay_nuenue}. Because it is the same dataset that is analyzed, limits are again less stringent than the expected sensitivity at energies around 1~TeV and the local signal significance reach modest values around $2.3\sigma$ ($p$-value of $\sim$ 1\%). 

\begin{figure*}
    \centering
    \begin{minipage}{.49\textwidth}
    \includegraphics[width=\linewidth]{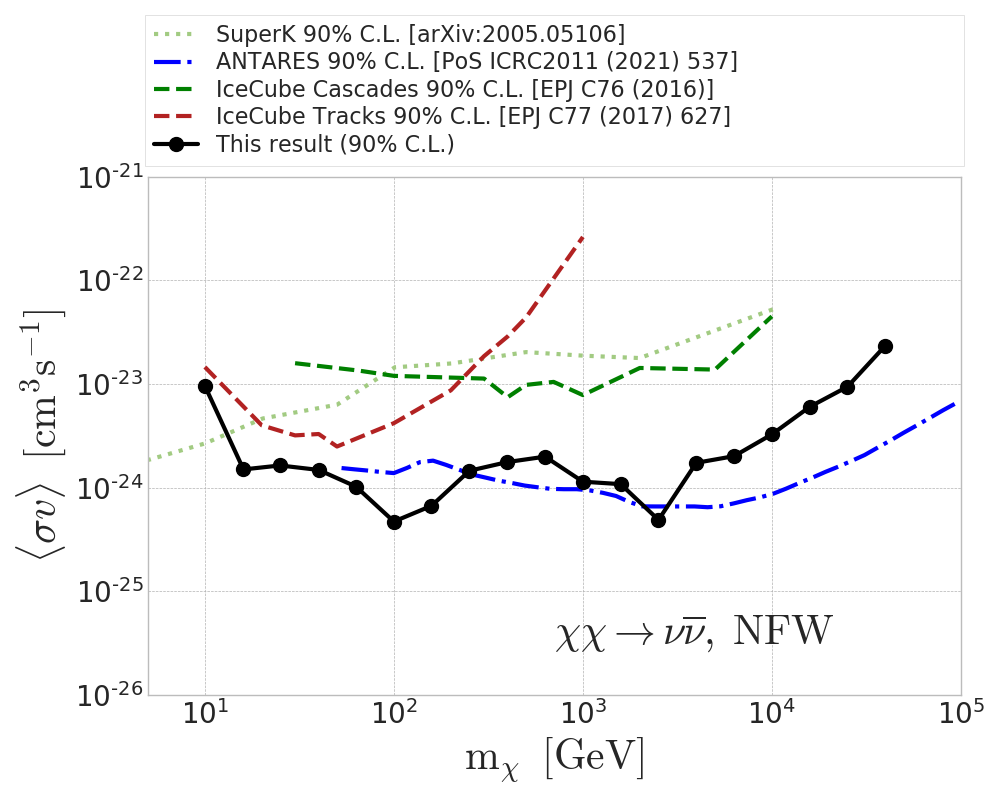}
    \end{minipage}
    \hfill
    \begin{minipage}{.49\textwidth}
    \includegraphics[width=\linewidth]{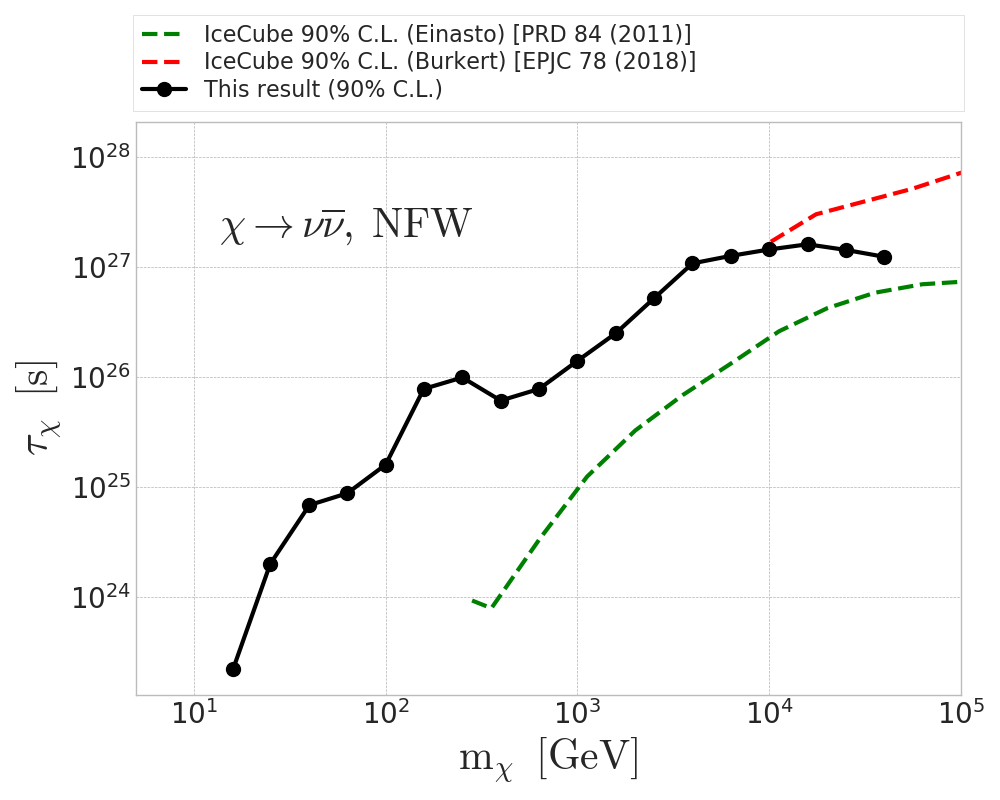}
    \end{minipage}
    \caption{Left: Limits on the thermally averaged cross-section for the average $\nu\overline{\nu}$ channels compared to previous IceCube results~\cite{IC86_GCWIMP_Event, IC86_GCWIMP} as well as Super-Kamiokande~\cite{SuperK} and ANTARES~\cite{ANTARES_POS}. Right: Limits on the decaying lifetime for the average $\nu\overline{\nu}$ channels compared also to previous IceCube limits~\cite{GalacticHalo,decayingDM}.} 
    \label{fig:results_other_experiments_nue}

\end{figure*}

Fig.~\ref{fig:results_other_experiments_nue} shows the results of dark matter annihilating (left) and decaying (right) to neutrinos (the average limit over the three neutrino flavor channels) in comparison with other neutrino experiments.

In the annihilation mode there is a notable improvement of $\sim \mathcal{O}(10)$ for masses above 100 GeV when compared to IceCube's previous results using a similar event selection and one year of IceCube data~\cite{IC86_GCWIMP_Event}. This significant improvement is realized by considering both the angular and energy information of the neutrino events together with additional years of data. 

There is still room for further improvement of these limits in the near future. First of all, more years of data are available and will improve IceCube's sensivity to dark matter. In addition, recent technical improvements within the collaboration, such as better cascade energy and directional reconstructions using Deep Neural Networks~\cite{DNN} together with a better understanding and modeling of the ice properties and calibration of the photo-detector response functions, will improve the energy resolution making it more sensitive to dark matter monochromatic signatures.

Note also that even if the $\sim 10^{-24}$ sensitivity reached on ${\langle \sigma v\rangle}$ in Fig.~\ref{fig:results_annihilation_nuenue} is a factor $\sim 30$ larger than 
the annihilation cross section at the time of DM freeze-out in thermal DM frameworks, the Sommerfeld enhancement effect 
 can largely boost this cross section into neutrinos today in the Galactic Center~\cite{ElAisati:2017ppn}. As a result neutrino 
telescopes are already testing today thermal scenarios where DM annihilates for a large part into neutrinos. This basically only requires that the mediator 
through which DM annihilate into neutrinos is sufficiently lighter than the DM particle. This holds for DM masses above few TeV if the mediator
is an electroweak gauge boson or below if the mediator is a new lighter particle beyond the Standard Model.

%%%%%%%%%%%%%%%%%%%%%%%%%%%%%%%%%%%%%%%%%%%%%%%%%%%%%%%%%%%%%%%%%%%%%%%%%%%%%%%%%
\section{Results for the secondary neutrino channels}\label{sec:results2}
For annihilation and decay channels proceeding into a pair of Standard Model charged particles, leading to a continuous energy spectrum of secondary neutrinos, the energy information
of the event is less crucial than for a monochromatic line. However, using the energy information of the events still leads to an improvement of the sensitivity.

Using the same data samples as for the neutrino line searches, no clear deviation from the background hypothesis is observed with any of the DM annihilation and decay channels tested. The mild excess of events observed in the $\nu_\alpha\overline{\nu}_\alpha$ channels is also observed for these channels, although at slightly higher masses (this is specially true for the $b\overline{b}$ channel). The most significant excess among all our studied signals shows up for the annihilation into $\tau^+\tau^-$ final states at $\sim$ 1.5 TeV with the Burkert profile, which yields a pre-trial significance of 0.03. 

However, correcting for the number of trials $-$due to the different channels, masses, and DM profiles analyzed$-$, by generating background pseudo-samples and repeating the analysis, lowers the significance to $\sim 38\%$, which is well compatible with the background expectations. 

\begin{figure*}
    \centering
    \begin{minipage}{.49\textwidth}
    \includegraphics[width=\linewidth]{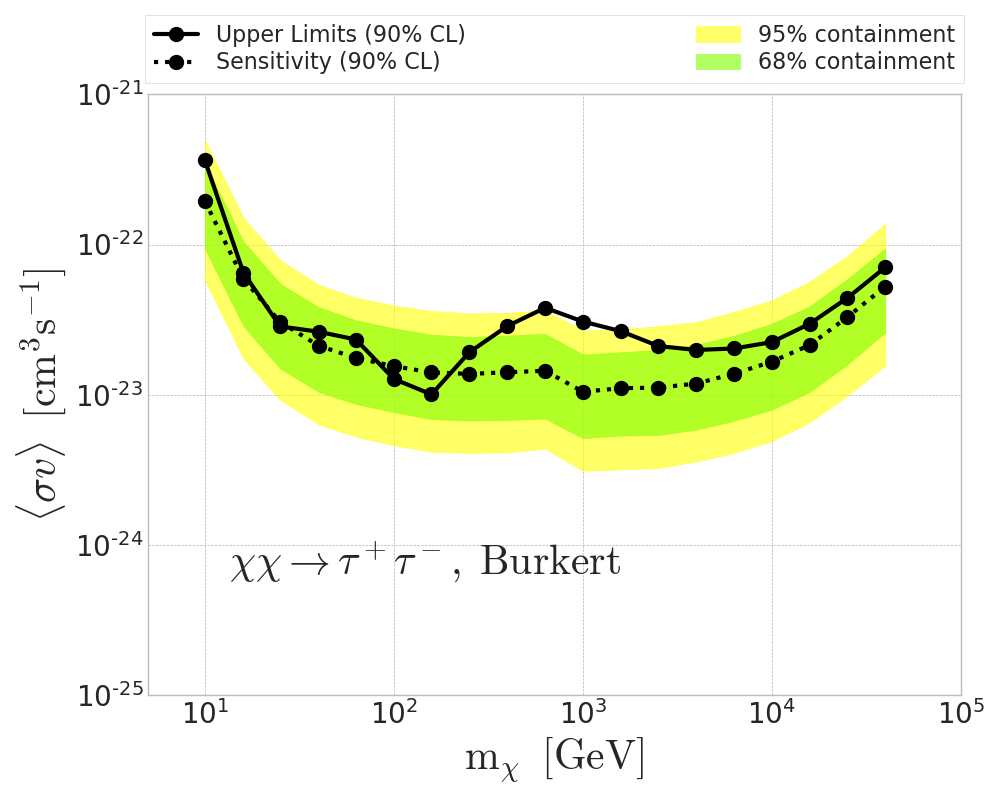}
    \end{minipage}
    \hfill
    \begin{minipage}{.49\textwidth}
    \includegraphics[width=\linewidth]{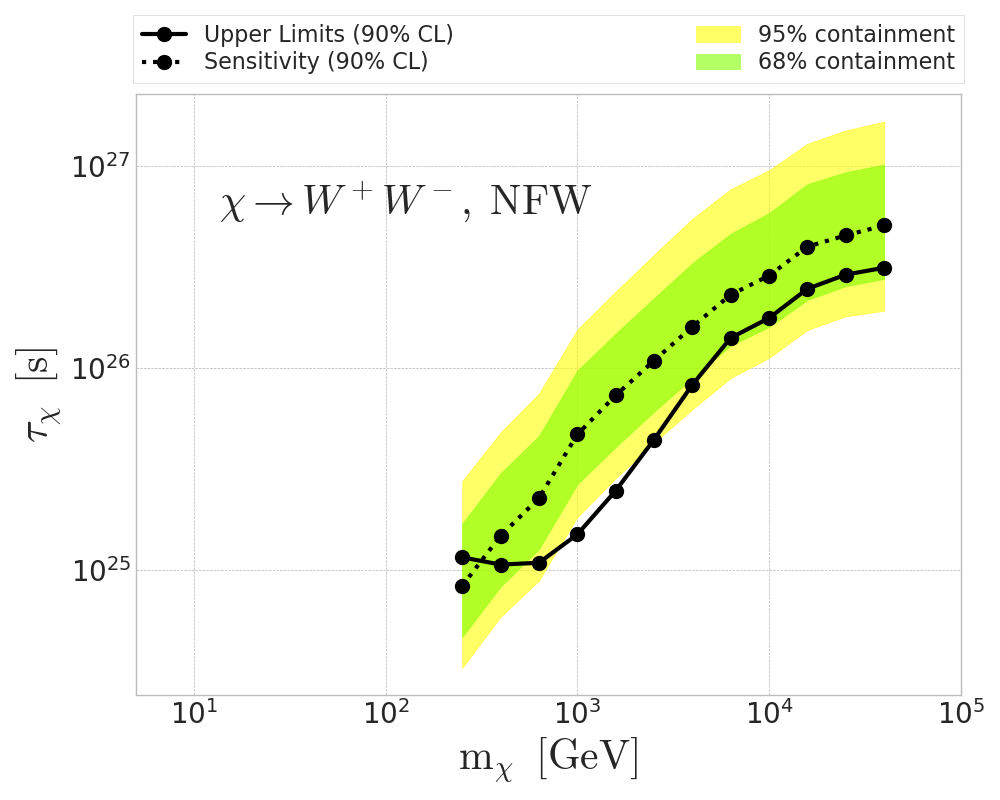}
    \end{minipage}
    
    \caption{Left: Same as Fig.~\ref{fig:results_annihilation_nuenue}, but for $\tau^+\tau^-$ annihilation channel and Burkert profile. Right: Lower limits, and sensitivity, on the decaying lifetime for $W^+ W^-$ and NFW profile.} 
    \label{fig:results_other_channelsA}
\end{figure*}

In Fig.~\ref{fig:results_other_channelsA} we show the results for the annihilation into $\tau^+\tau^-$ channel and the Burkert profile, as well as the results for dark matter decay into $W^+ W^-$ with the NFW profile. 
 
The results of all the channels and profiles tested can be found in the Appendix, together with the plots summarizing all the results.

\begin{figure*}
    \centering
    \begin{minipage}{.49\textwidth}
    \includegraphics[width=\linewidth]{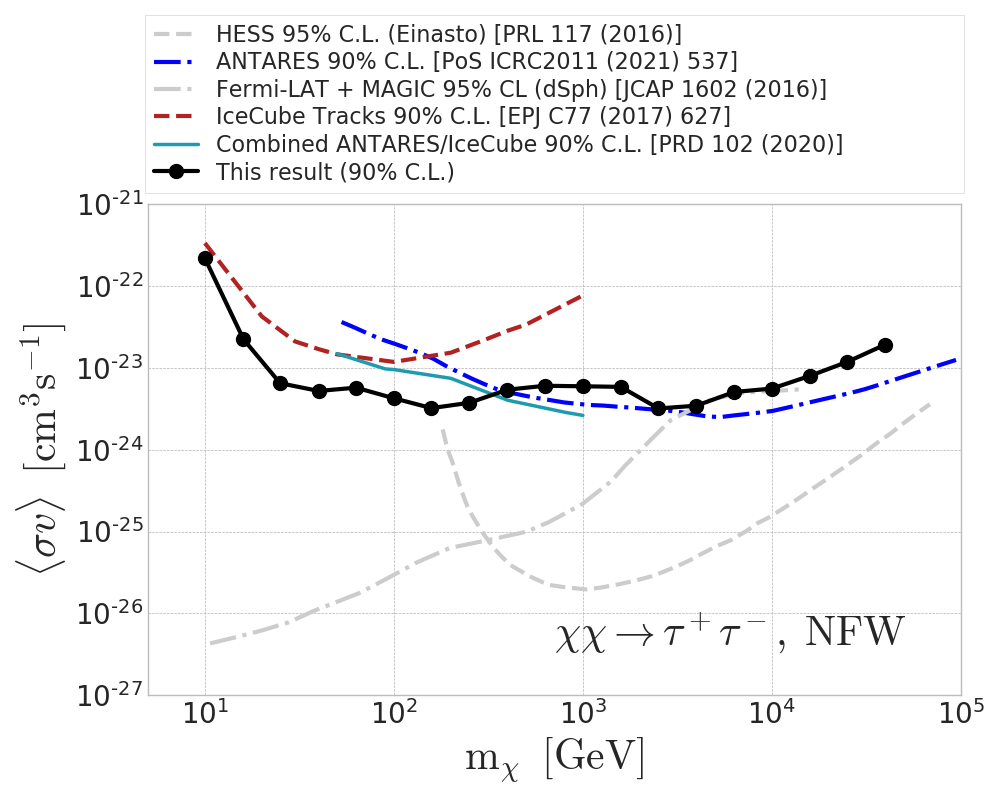}
    \end{minipage}
    \hfill
    \begin{minipage}{.49\textwidth}
    \includegraphics[width=\linewidth]{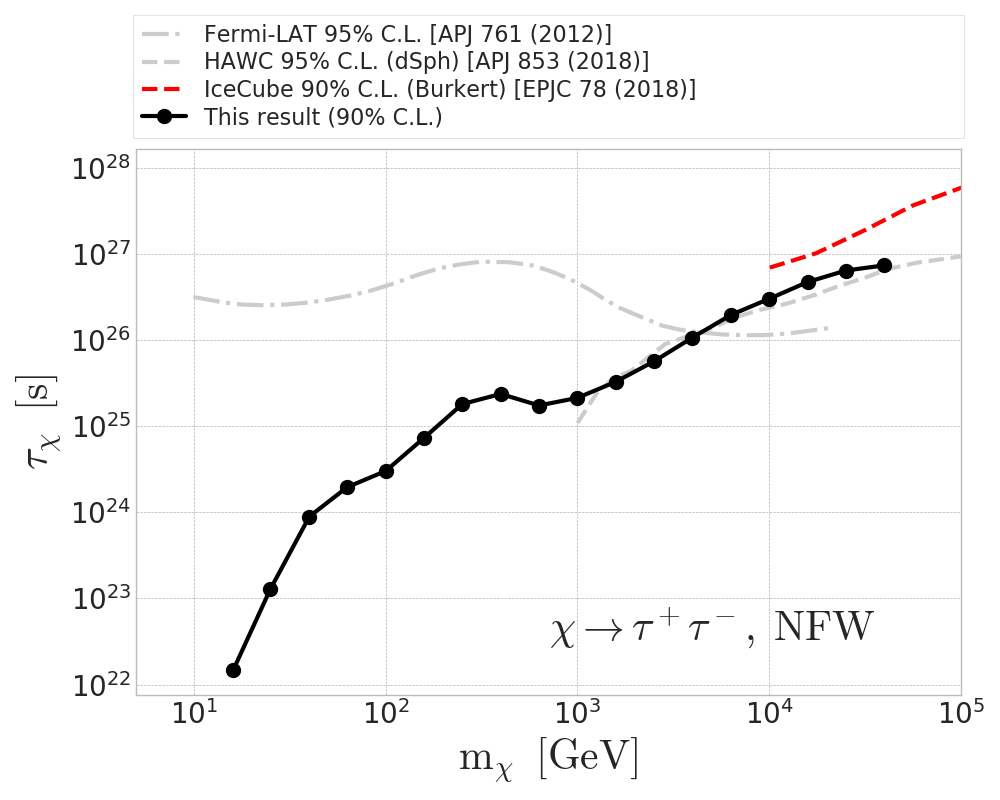}
    \end{minipage}
    \caption{
    Left: Upper limits on the thermally averaged cross-section for the $\tau^+\tau^-$ channel compared to other neutrino detectors such as previous IceCube results~\cite{IC86_GCWIMP_Event, IC86_GCWIMP} and ANTARES~\cite{ANTARES_POS}, and the gamma-ray telescope H.E.S.S~\cite{HESS} and Fermi~\cite{Fermi-MAGIC}. Right: Lower limits on the dark matter decay lifetime for the $\tau^+\tau^-$ channel compared also to previous IceCube limits~\cite{GalacticHalo,decayingDM} and HAWC~\cite{HAWC}.}
    \label{fig:results_other_experiments}
\end{figure*}

Fig.~\ref{fig:results_other_experiments} shows the results obtained for the $\tau^+ \tau^-$ channel in comparison with other neutrino and gamma-ray experiments. For the charged particle channels the portion of energy that goes into $\gamma$'s or $e^{\pm}$, quickly producing $\gamma$-rays, is in general large~\cite{Cirelli}. Thus, combined with the fact that neutrinos have lower detection cross sections, gamma-ray detectors are in general more sensitive to these channels than neutrino telescopes.

 It is also interesting to compare the limits obtained above on neutrino pair production by neutrino telescopes with the limits obtained by gamma-ray telescopes on charged lepton pair production. This comparison is interesting because in  many models the associated annihilation cross sections are predicted to be basically equal due to $SU(2)_L$ gauge invariance, even if it is known that there also exist other models where the charged pair production is way more suppressed than the neutrino pair production, see \cite{ElAisati:2017ppn}.

As can be seen in Fig.~\ref{fig:results_other_experiments}, gamma-ray telescopes are still more sensitive on the DM production of $\tau^+ \tau^-$ than neutrino telescopes on the production of the $\nu\bar{\nu}$ channel (see Fig.~\ref{fig:results_other_experiments_nue}) by one order of magnitude at masses below few tens of TeV. On the other hand, neutrino telescope limits on the neutrino channel are comparable to gamma-ray limits on the $\mu^+ \mu^-$ and $e^+ e^-$ channels for masses above a few TeV \cite{MAGIC:2016xys,John:2021ugy,Bergstrom:2013jra}.

\section{Conclusion}
In this work we showed the results of the first neutrino telescope dedicated search for neutrino lines, using both the spatial and energy information of the neutrino events. The event selections, both the low-energy and the high-energy, are based on a five-year cascade event IceCube/DeepCore data sample~\cite{IC86_GCWIMP_Event}.
No evidence of dark matter signature was found and new upper limits (lower limits) were set on the annihilation cross-section (decay lifetime).
The results constitutes a large improvement with respect to previous analyses of the order of one order of magnitude, except for DM masses around 1 TeV where the improvement is less significant due to a mild excess of neutrino events causing weaker DM constraints as compared to the expected sensitivity. The same analyses provides competitive limits for DM annihilation and decay into charged particles. More available data as well as new advancements in cascade reconstructions and MC will be able to improve these limits in the near future.

\label{sec:conclusion}

%\section*{Acknowledgements}
\begin{acknowledgments}

The IceCube collaboration acknowledges the significant contributions to this manuscript from J. A. Aguilar, M. Gustafsson, and T. Hambye.
USA -- U.S. National Science Foundation-Office of Polar Programs,
U.S. National Science Foundation-Physics Division,
Wisconsin Alumni Research Foundation,
Center for High Throughput Computing (CHTC) at the University of Wisconsin-Madison,
Open Science Grid (OSG),
Extreme Science and Engineering Discovery Environment (XSEDE),
U.S. Department of Energy-National Energy Research Scientific Computing Center,
Particle astrophysics research computing center at the University of Maryland,
Institute for Cyber-Enabled Research at Michigan State University,
and Astroparticle physics computational facility at Marquette University;
Belgium -- Funds for Scientific Research (FRS-FNRS and FWO),
FWO Odysseus and Big Science programmes,
and Belgian Federal Science Policy Office (Belspo);
Germany -- Bundesministerium f\"ur Bildung und Forschung (BMBF),
Deutsche Forschungsgemeinschaft (DFG),
Helmholtz Alliance for Astroparticle Physics (HAP),
Initiative and Networking Fund of the Helmholtz Association,
Deutsches Elektronen Synchrotron (DESY),
and High Performance Computing cluster of the RWTH Aachen;
Sweden -- Swedish Research Council,
Swedish Polar Research Secretariat,
Swedish National Infrastructure for Computing (SNIC),
and Knut and Alice Wallenberg Foundation;
Australia -- Australian Research Council;
Canada -- Natural Sciences and Engineering Research Council of Canada,
Calcul Qu\'ebec, Compute Ontario, Canada Foundation for Innovation, WestGrid, and Compute Canada;
Denmark -- Villum Fonden, Danish National Research Foundation (DNRF);
New Zealand -- Marsden Fund;
Japan -- Japan Society for Promotion of Science (JSPS)
and Institute for Global Prominent Research (IGPR) of Chiba University;
Korea -- National Research Foundation of Korea (NRF);
Switzerland -- Swiss National Science Foundation (SNSF). This particular project has been further supported in Belgium by the “Probing dark matter with
neutrinos” ULB-ARC convention, by the Excellence of Science (EoS) project No. 30820817 - be.h “The H boson gateway to physics beyond the
Standard Model”, by the FRIA and IISN (No. 4.4503.15)

\end{acknowledgments}

%\appendix

\section{Appendix}
\label{sec:appendix}
\begin{figure*}
    \centering
    \begin{minipage}{.49\textwidth}
    \includegraphics[width=\linewidth]{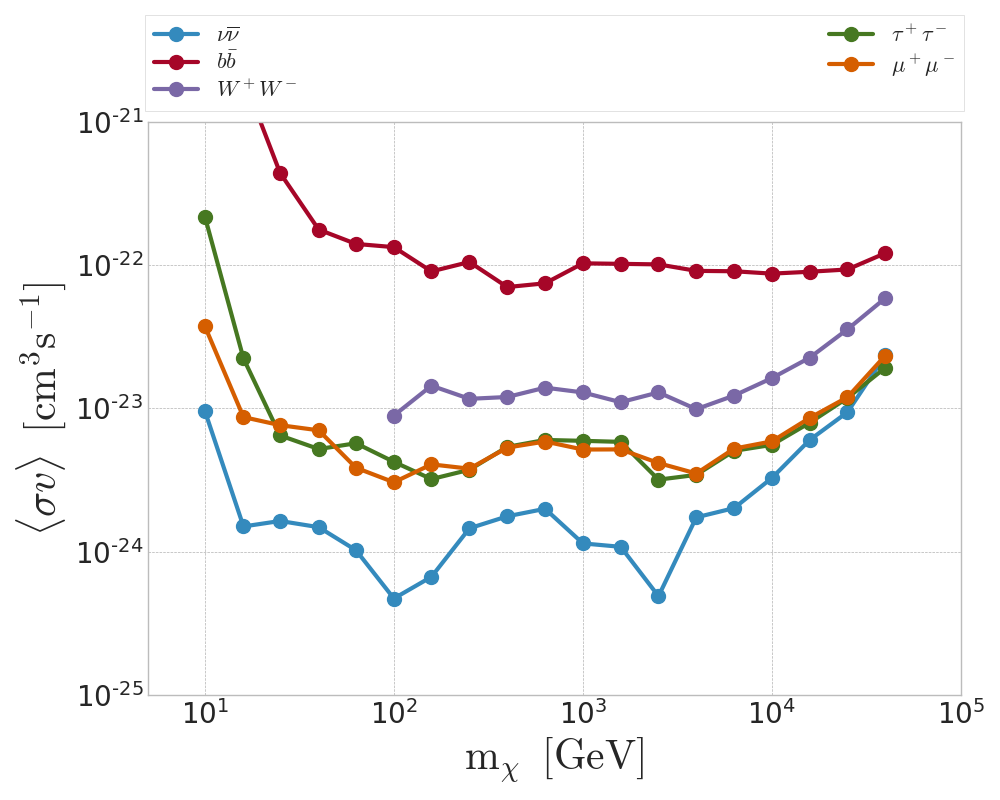}
    \end{minipage}
    \hfill
    \begin{minipage}{.49\textwidth}
    \includegraphics[width=\linewidth]{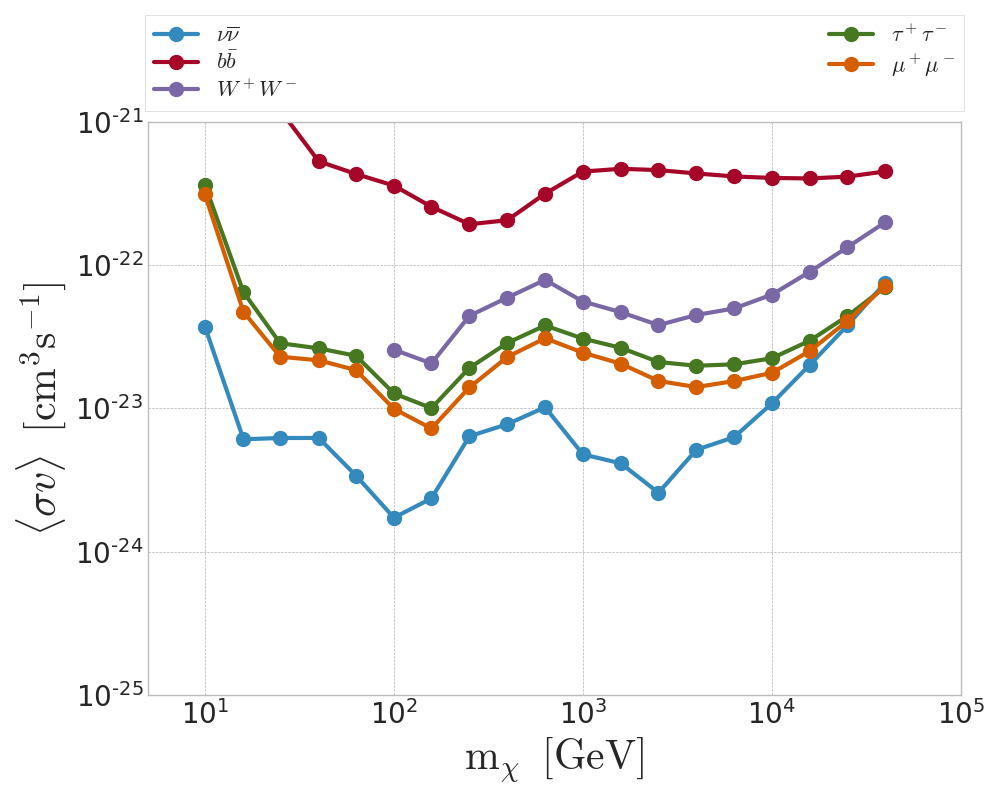}
    \end{minipage}
    \begin{minipage}{.49\textwidth}
    \includegraphics[width=\linewidth]{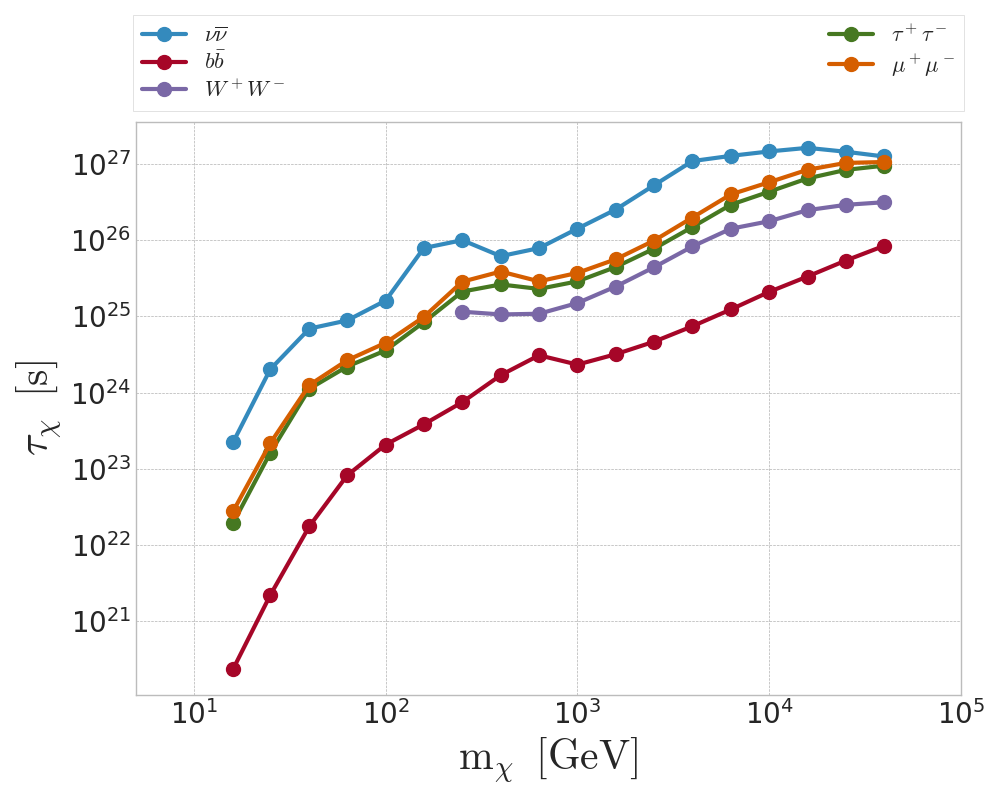}
    \end{minipage}
    \hfill
    \begin{minipage}{.49\textwidth}
    \includegraphics[width=\linewidth]{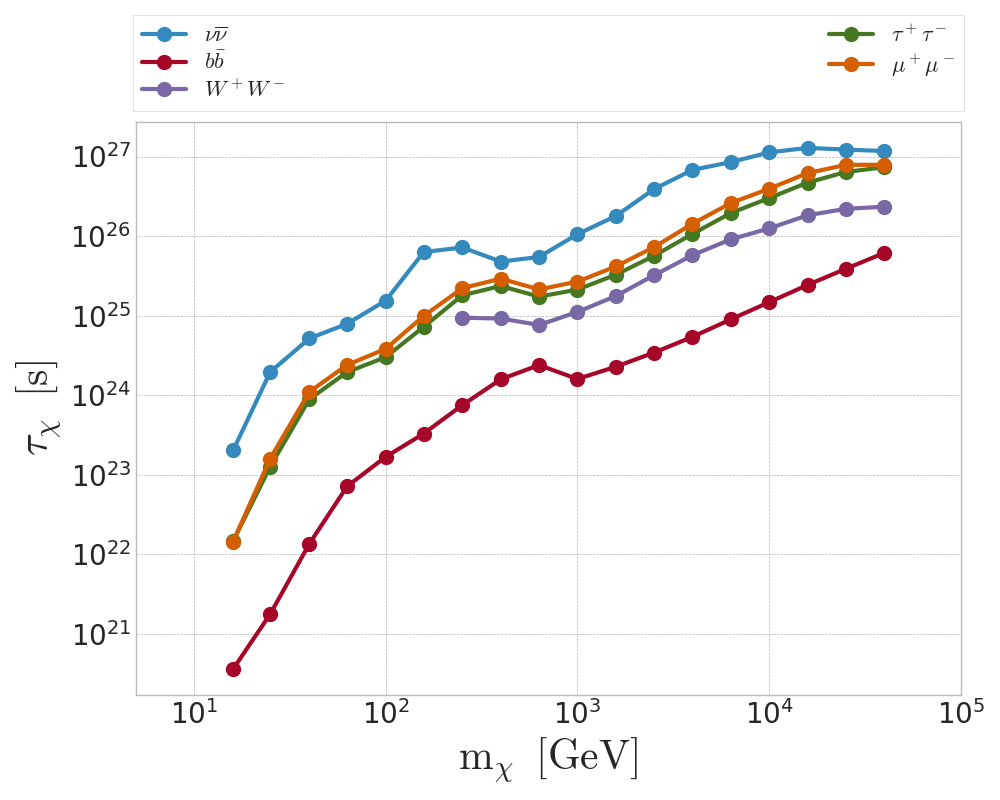}
    \end{minipage}
    \caption{Top: Upper limits for all channels in thermally averaged cross-section as function of the dark matter mass for the NFW profile (left) and Burkert profile (right). Bottom: Lower-limits for all channels in dark matter lifetime as function of the dark matter mass for the NFW profile (left) and Burkert profile (right).} 
    \label{fig:results_other_channels}
\end{figure*}

Fig.~\ref{fig:results_other_channels} shows the limits obtained for an annihilation and decay case for all the annihilation and decay channels tested: $\nu\bar{\nu}$, $\mu^+\mu^-$, $\tau^+\tau^-$, $b\bar{b}$  and $W^+W^-$ channels. The neutrino-line channel is the average from all the three neutrino flavors. This plots summarizes the information that can be found on the following tables. 

\begin{center}
\begin{table*}
\begin{tabular}{|r| c c c | c c c |  c c c |  c c c | }
\toprule
 & \multicolumn{6}{c|}{{\bf Annihilation}} & \multicolumn{6}{c|}{{\bf Decay}} \\
 & \multicolumn{3}{c|}{{\bf NFW }} &  \multicolumn{3}{c|}{{\bf Burkert}} & \multicolumn{3}{c|}{{\bf NFW}} &  \multicolumn{3}{c|}{\bf{Burkert}} \\
{$m_\chi$} &  \multirow{2}{*}{$\hat{n}_s$} & {$\langle \sigma v \rangle^{90\%}_{u.l.} $} & \multirow{2}{*}{$z$-score} &  \multirow{2}{*}{$\hat{n}_s$} & {$\langle \sigma v \rangle^{90\%}_{u.l.}$}   & \multirow{2}{*}{$z$-score} &  \multirow{2}{*}{$\hat{n}_s$} & {$\tau^{90\%}_{l.l.}$} & \multirow{2}{*}{$z$-score} &  \multirow{2}{*}{$\hat{n}_s$} & {$\tau^{90\%}_{l.l.}$} & \multirow{2}{*}{$z$-score} \\ 
${\rm [GeV]}$ &   & {$10^{-24} {\rm \;[cm^2]}$} &   &   & {$10^{-24} {\rm \;[cm^2]}$} &  &  &  {$10^{24} {\rm \;[s]}$} &   &    & {$10^{24} {\rm \;[s]}$} & \\
\midrule
10 & 51.99 & 15.01 & 0.50 & 8.90 & 43.02 & 0.04 & - & - & - & - & - & - \\
16 & 0.00 & 2.26 & 0.00 & 0.00 & 6.65 & 0.00 & 24.36 & 0.23 & 0.10 & 0.01 & 0.24 & 0.00 \\
25 & 68.06 & 2.14 & 0.41 & 2.44 & 5.81 & 0.01 & 0.00 & 1.90 & 0.00 & 0.00 & 2.00 & 0.00 \\
40 & 132.26 & 1.92 & 0.79 & 142.46 & 6.16 & 0.45 & 0.00 & 7.26 & 0.00 & 0.21 & 4.63 & 0.00 \\
63 & 48.95 & 1.21 & 0.36 & 0.00 & 3.06 & 0.00 & 203.34 & 8.57 & 0.41 & 53.77 & 8.04 & 0.09 \\
100 & 0.00 & 0.42 & 0.00 & 0.00 & 1.76 & 0.00 & 207.92 & 16.20 & 0.45 & 131.84 & 13.80 & 0.22 \\
158 & 0.00 & 0.61 & 0.00 & 0.00 & 2.21 & 0.00 & 0.00 & 79.06 & 0.00 & 0.00 & 71.83 & 0.00 \\
251 & 68.15 & 1.41 & 1.41 & 162.78 & 6.43 & 1.58 & 0.00 & 97.71 & 0.00 & 0.00 & 71.84 & 0.00 \\
398 & 44.22 & 1.38 & 1.32 & 137.11 & 7.75 & 1.79 & 215.76 & 60.16 & 1.22 & 209.89 & 48.45 & 0.87 \\
631 & 20.81 & 1.35 & 0.77 & 123.15 & 10.30 & 1.94 & 232.55 & 79.23 & 1.69 & 320.05 & 54.84 & 1.73 \\
1000 & 58.75 & 1.13 & 1.27 & 116.62 & 4.73 & 1.22 & 397.88 & 139.26 & 2.29 & 491.00 & 103.89 & 2.15 \\
1585 & 32.94 & 0.96 & 0.95 & 67.87 & 4.11 & 0.95 & 277.51 & 248.91 & 1.70 & 369.96 & 177.36 & 1.73 \\
2512 & 0.00 & 0.30 & 0.00 & 0.00 & 2.57 & 0.00 & 135.46 & 528.81 & 1.11 & 165.28 & 391.76 & 1.02 \\
3981 & 21.03 & 1.43 & 1.11 & 27.53 & 5.26 & 0.61 & 32.34 & 1079.73 & 0.31 & 72.36 & 683.29 & 0.54 \\
6310 & 21.03 & 1.79 & 1.43 & 21.62 & 6.36 & 0.64 & 29.86 & 1310.17 & 0.36 & 54.86 & 841.87 & 0.48 \\
10000 & 18.93 & 3.13 & 1.40 & 19.86 & 11.16 & 0.63 & 45.25 & 1463.26 & 0.83 & 46.25 & 1097.92 & 0.60 \\
15850 & 16.38 & 5.67 & 1.25 & 18.84 & 21.18 & 0.62 & 38.10 & 1584.29 & 0.78 & 31.83 & 1266.47 & 0.46 \\
25120 & 13.64 & 9.54 & 1.21 & 17.56 & 41.11 & 0.61 & 33.71 & 1379.79 & 0.74 & 26.28 & 1092.62 & 0.40 \\
39810 & 17.94 & 27.30 & 1.33 & 18.46 & 84.87 & 0.64 & 33.64 & 1150.55 & 0.79 & 26.17 & 922.08 & 0.43 \\
\bottomrule
\end{tabular}
\caption{Table with the results for the final state channel $\nu_{e}\overline{\nu}_{e}$ for both the annihilation and decaying mode and for both the NFW and Burkert profile. The best fit value on the number of signal evens $\hat{n}_s$ is shown together with the resulting upper limit in $\langle \sigma v \rangle^{90\%}_{u.l.}$ and lower limit on $\tau^{90\%}_{l.l.}$ along with the significance given in number of sigmas, $z$-score.}
\label{tb:nuenue}
\end{table*}
\end{center}

\begin{center}
\begin{table*}
\begin{tabular}{|r| c c c | c c c |  c c c |  c c c | }
\toprule
 & \multicolumn{6}{c|}{{\bf Annihilation}} & \multicolumn{6}{c|}{{\bf Decay}} \\
 & \multicolumn{3}{c|}{{\bf NFW }} &  \multicolumn{3}{c|}{{\bf Burkert}} & \multicolumn{3}{c|}{{\bf NFW}} &  \multicolumn{3}{c|}{\bf{Burkert}} \\
{$m_\chi$} &  \multirow{2}{*}{$\hat{n}_s$} & {$\langle \sigma v \rangle^{90\%}_{u.l.} $} & \multirow{2}{*}{$z$-score} &  \multirow{2}{*}{$\hat{n}_s$} & {$\langle \sigma v \rangle^{90\%}_{u.l.}$}   & \multirow{2}{*}{$z$-score} &  \multirow{2}{*}{$\hat{n}_s$} & {$\tau^{90\%}_{u.l.}$} & \multirow{2}{*}{$z$-score} &  \multirow{2}{*}{$\hat{n}_s$} & {$\tau^{90\%}_{u.l.}$} & \multirow{2}{*}{$z$-score} \\ 
${\rm [GeV]}$ &   & {$10^{-24} {\rm \;[cm^2]}$} &   &   & {$10^{-24} {\rm \;[cm^2]}$} &  &  &  {$10^{24} {\rm \;[s]}$} &   &    & {$10^{24} {\rm \;[s]}$} & \\
\midrule
10 & 27.65 & 8.60 & 0.47 & 0.00 & 37.50 & 0.00 & - & - & - & - & - & - \\
16 & 0.00 & 1.27 & 0.00 & 0.00 & 5.61 & 0.00 & 30.20 & 0.22 & 0.12 & 78.38 & 0.16 & 0.27 \\
25 & 0.00 & 1.24 & 0.00 & 39.96 & 6.37 & 0.13 & 0.00 & 2.19 & 0.00 & 0.73 & 1.46 & 0.00 \\
40 & 0.00 & 1.14 & 0.00 & 131.83 & 6.00 & 0.42 & 0.00 & 6.78 & 0.00 & 0.00 & 5.53 & 0.00 \\
63 & 0.01 & 1.04 & 0.00 & 0.00 & 3.57 & 0.00 & 152.05 & 9.11 & 0.31 & 37.92 & 8.13 & 0.06 \\
100 & 0.00 & 0.44 & 0.00 & 0.00 & 1.74 & 0.00 & 218.75 & 15.92 & 0.47 & 0.00 & 17.62 & 0.00 \\
158 & 0.00 & 0.58 & 0.00 & 0.00 & 2.57 & 0.00 & 0.00 & 73.24 & 0.00 & 0.00 & 59.21 & 0.00 \\
251 & 70.75 & 1.40 & 1.27 & 159.23 & 6.34 & 1.55 & 0.00 & 99.24 & 0.00 & 0.00 & 78.70 & 0.00 \\
398 & 68.09 & 2.09 & 1.69 & 141.16 & 7.87 & 1.84 & 194.77 & 62.38 & 1.07 & 205.03 & 48.75 & 0.84 \\
631 & 36.75 & 2.20 & 1.09 & 120.84 & 10.15 & 1.89 & 241.20 & 77.63 & 1.74 & 323.28 & 54.62 & 1.75 \\
1000 & 65.46 & 1.24 & 1.24 & 118.37 & 4.75 & 1.23 & 397.15 & 139.88 & 2.30 & 468.35 & 107.05 & 2.05 \\
1585 & 23.72 & 1.07 & 0.59 & 69.60 & 4.13 & 0.97 & 270.54 & 253.60 & 1.66 & 370.16 & 177.54 & 1.72 \\
2512 & 0.01 & 0.58 & 0.00 & 0.02 & 2.58 & 0.00 & 142.02 & 518.33 & 1.15 & 171.01 & 387.55 & 1.05 \\
3981 & 24.79 & 1.85 & 1.13 & 25.04 & 5.02 & 0.55 & 35.16 & 1089.38 & 0.34 & 77.10 & 681.11 & 0.57 \\
6310 & 18.17 & 2.01 & 1.14 & 22.41 & 6.25 & 0.64 & 38.00 & 1237.17 & 0.44 & 56.26 & 843.13 & 0.48 \\
10000 & 24.25 & 3.69 & 1.60 & 20.93 & 10.66 & 0.65 & 49.90 & 1423.89 & 0.87 & 45.37 & 1128.16 & 0.57 \\
15850 & 22.15 & 6.45 & 1.48 & 19.59 & 19.46 & 0.63 & 38.22 & 1654.79 & 0.76 & 33.78 & 1304.98 & 0.48 \\
25120 & 14.89 & 9.57 & 1.28 & 18.57 & 36.17 & 0.62 & 34.14 & 1494.90 & 0.73 & 28.97 & 1151.67 & 0.43 \\
39810 & 17.03 & 20.15 & 1.46 & 18.17 & 70.14 & 0.62 & 32.56 & 1314.54 & 0.74 & 0.00 & 1576.34 & 0.00 \\
\bottomrule
\end{tabular}
\caption{Same as table~\ref{tb:nuenue} for the $\nu_{\mu}\overline{\nu}_{\mu}$ channel.
Since we assumed a democratic neutrino flavor compositions of the signals at the detector, any differences among the tables are dominantly due to statistical fluctuations between the signals' generated {\it pdf's}.}
\label{tb:numunumu}
\end{table*}
\end{center}

\begin{center}
\begin{table*}
\begin{tabular}{|r| c c c | c c c |  c c c |  c c c | }
\toprule
 & \multicolumn{6}{c|}{{\bf Annihilation}} & \multicolumn{6}{c|}{{\bf Decay}} \\
 & \multicolumn{3}{c|}{{\bf NFW }} &  \multicolumn{3}{c|}{{\bf Burkert}} & \multicolumn{3}{c|}{{\bf NFW}} &  \multicolumn{3}{c|}{\bf{Burkert}} \\
{$m_\chi$} &  \multirow{2}{*}{$\hat{n}_s$} & {$\langle \sigma v \rangle^{90\%}_{u.l.} $} & \multirow{2}{*}{$z$-score} &  \multirow{2}{*}{$\hat{n}_s$} & {$\langle \sigma v \rangle^{90\%}_{u.l.}$}   & \multirow{2}{*}{$z$-score} &  \multirow{2}{*}{$\hat{n}_s$} & {$\tau^{90\%}_{u.l.}$} & \multirow{2}{*}{$z$-score} &  \multirow{2}{*}{$\hat{n}_s$} & {$\tau^{90\%}_{u.l.}$} & \multirow{2}{*}{$z$-score} \\ 
${\rm [GeV]}$ &   & {$10^{-24} {\rm \;[cm^2]}$} &   &   & {$10^{-24} {\rm \;[cm^2]}$} &  &  &  {$10^{24} {\rm \;[s]}$} &   &    & {$10^{24} {\rm \;[s]}$} & \\
\midrule
10 & 0.01 & 5.17 & 0.00 & 0.00 & 31.36 & 0.00 & - & - & - & - & - & - \\
16 & 0.00 & 0.98 & 0.00 & 0.00 & 6.04 & 0.00 & 31.15 & 0.22 & 0.12 & 6.49 & 0.20 & 0.02 \\
25 & 17.12 & 1.55 & 0.12 & 48.89 & 6.49 & 0.16 & 0.00 & 1.96 & 0.00 & 0.00 & 2.42 & 0.00 \\
40 & 31.65 & 1.40 & 0.20 & 173.99 & 6.53 & 0.55 & 0.00 & 6.67 & 0.00 & 0.00 & 5.34 & 0.00 \\
63 & 0.00 & 0.83 & 0.00 & 0.00 & 3.58 & 0.00 & 184.14 & 8.83 & 0.38 & 95.94 & 7.59 & 0.15 \\
100 & 0.00 & 0.56 & 0.00 & 0.00 & 1.69 & 0.00 & 206.01 & 16.19 & 0.45 & 66.11 & 14.91 & 0.11 \\
158 & 0.00 & 0.81 & 0.00 & 0.00 & 2.30 & 0.00 & 0.00 & 83.23 & 0.00 & 0.00 & 57.91 & 0.00 \\
251 & 86.37 & 1.57 & 1.49 & 162.47 & 6.43 & 1.58 & 0.00 & 102.25 & 0.00 & 0.14 & 65.23 & 0.00 \\
398 & 55.98 & 1.84 & 1.48 & 136.87 & 7.70 & 1.79 & 202.46 & 62.01 & 1.14 & 235.25 & 46.44 & 0.98 \\
631 & 46.98 & 2.43 & 1.44 & 119.94 & 10.15 & 1.87 & 235.96 & 78.56 & 1.71 & 325.58 & 54.40 & 1.76 \\
1000 & 46.34 & 1.06 & 0.87 & 125.21 & 4.88 & 1.30 & 377.12 & 144.53 & 2.18 & 475.38 & 106.10 & 2.08 \\
1585 & 33.36 & 1.21 & 0.84 & 72.25 & 4.20 & 1.01 & 272.23 & 252.43 & 1.67 & 358.44 & 181.06 & 1.67 \\
2512 & 0.00 & 0.59 & 0.00 & 0.02 & 2.61 & 0.00 & 138.05 & 525.37 & 1.12 & 168.37 & 390.51 & 1.03 \\
3981 & 26.26 & 1.95 & 1.15 & 27.16 & 5.16 & 0.59 & 35.13 & 1088.77 & 0.34 & 79.86 & 669.90 & 0.59 \\
6310 & 22.50 & 2.26 & 1.38 & 22.59 & 6.31 & 0.64 & 36.43 & 1265.33 & 0.43 & 50.88 & 865.59 & 0.44 \\
10000 & 18.16 & 3.00 & 1.35 & 21.13 & 10.74 & 0.66 & 47.30 & 1470.41 & 0.84 & 44.03 & 1148.93 & 0.56 \\
15850 & 19.28 & 5.98 & 1.29 & 19.94 & 19.73 & 0.63 & 40.21 & 1618.85 & 0.79 & 33.85 & 1282.60 & 0.47 \\
25120 & 13.84 & 9.17 & 1.16 & 19.80 & 36.90 & 0.66 & 36.48 & 1442.87 & 0.77 & 0.00 & 1415.95 & 0.00 \\
39810 & 21.64 & 22.87 & 1.75 & 19.93 & 71.50 & 0.66 & 34.10 & 1275.83 & 0.76 & 13.62 & 1018.97 & 0.18 \\
\bottomrule
\end{tabular}
\caption{Same as table~\ref{tb:nuenue} for the $\nu_{\tau}\overline{\nu}_{\tau}$ channel. Since we assumed a democratic neutrino flavor compositions of the signals at the detector, any differences among the tables are dominantly due to statistical fluctuations between the signals' generated {\it pdf's}.}
\label{tb:nutaunutau}
\end{table*}
\end{center}

\begin{center}
\begin{table*}
\begin{tabular}{|r| c c c | c c c |  c c c |  c c c | }
\toprule
 & \multicolumn{6}{c|}{{\bf Annihilation}} & \multicolumn{6}{c|}{{\bf Decay}} \\
 & \multicolumn{3}{c|}{{\bf NFW }} &  \multicolumn{3}{c|}{{\bf Burkert}} & \multicolumn{3}{c|}{{\bf NFW}} &  \multicolumn{3}{c|}{\bf{Burkert}} \\
{$m_\chi$} &  \multirow{2}{*}{$\hat{n}_s$} & {$\langle \sigma v \rangle^{90\%}_{u.l.} $} & \multirow{2}{*}{$z$-score} &  \multirow{2}{*}{$\hat{n}_s$} & {$\langle \sigma v \rangle^{90\%}_{u.l.}$}   & \multirow{2}{*}{$z$-score} &  \multirow{2}{*}{$\hat{n}_s$} & {$\tau^{90\%}_{u.l.}$} & \multirow{2}{*}{$z$-score} &  \multirow{2}{*}{$\hat{n}_s$} & {$\tau^{90\%}_{u.l.}$} & \multirow{2}{*}{$z$-score} \\ 
${\rm [GeV]}$ &   & {$10^{-24} {\rm \;[cm^2]}$} &   &   & {$10^{-24} {\rm \;[cm^2]}$} &  &  &  {$10^{24} {\rm \;[s]}$} &   &    & {$10^{24} {\rm \;[s]}$} & \\
\midrule
10 & 41.34 & 9340.72 & 0.53 & 25.05 & 22335.36 & 0.19 & - & - & - & - & - & - \\
16 & 63.32 & 2029.91 & 0.57 & 0.00 & 4620.84 & 0.00 & 80.40 & 0.00 & 0.46 & 0.00 & 0.00 & 0.00 \\
25 & 4.98 & 439.28 & 0.03 & 0.00 & 1208.43 & 0.00 & 31.76 & 0.00 & 0.12 & 20.79 & 0.00 & 0.07 \\
40 & 0.00 & 177.03 & 0.00 & 0.00 & 529.93 & 0.00 & 0.31 & 0.02 & 0.00 & 0.00 & 0.01 & 0.00 \\
63 & 34.33 & 140.57 & 0.18 & 0.04 & 432.09 & 0.00 & 0.10 & 0.08 & 0.00 & 0.00 & 0.07 & 0.00 \\
100 & 127.42 & 133.51 & 0.60 & 45.81 & 358.84 & 0.12 & 0.00 & 0.21 & 0.00 & 0.00 & 0.17 & 0.00 \\
158 & 67.91 & 90.43 & 0.32 & 0.00 & 255.22 & 0.00 & 0.08 & 0.38 & 0.00 & 0.00 & 0.33 & 0.00 \\
251 & 99.51 & 105.18 & 0.47 & 0.00 & 192.89 & 0.00 & 16.88 & 0.75 & 0.03 & 0.00 & 0.75 & 0.00 \\
398 & 56.28 & 70.32 & 0.32 & 0.00 & 205.67 & 0.00 & 0.00 & 1.69 & 0.00 & 0.00 & 1.58 & 0.00 \\
631 & 109.59 & 74.69 & 0.76 & 196.62 & 315.35 & 0.64 & 0.14 & 3.08 & 0.00 & 0.00 & 2.40 & 0.00 \\
1000 & 207.59 & 102.94 & 2.00 & 444.50 & 449.47 & 2.07 & 366.60 & 2.33 & 1.12 & 525.96 & 1.60 & 1.22 \\
1585 & 233.26 & 102.09 & 2.31 & 516.36 & 469.91 & 2.42 & 623.24 & 3.19 & 1.89 & 821.49 & 2.28 & 1.89 \\
2512 & 229.47 & 101.20 & 2.37 & 498.33 & 460.13 & 2.44 & 802.91 & 4.65 & 2.45 & 1003.51 & 3.44 & 2.31 \\
3981 & 182.86 & 91.19 & 2.04 & 425.56 & 436.13 & 2.26 & 820.11 & 7.41 & 2.58 & 1044.57 & 5.41 & 2.49 \\
6310 & 156.50 & 90.63 & 1.94 & 341.57 & 415.65 & 2.02 & 737.14 & 12.30 & 2.47 & 936.61 & 8.99 & 2.38 \\
10000 & 122.11 & 87.24 & 1.78 & 263.93 & 405.73 & 1.79 & 591.45 & 20.92 & 2.19 & 789.43 & 14.76 & 2.19 \\
15850 & 98.44 & 89.93 & 1.73 & 197.21 & 402.38 & 1.58 & 486.66 & 33.27 & 2.01 & 618.35 & 24.31 & 1.93 \\
25120 & 71.63 & 93.29 & 1.50 & 144.07 & 414.40 & 1.36 & 369.04 & 53.94 & 1.79 & 475.30 & 38.83 & 1.72 \\
39810 & 63.13 & 121.20 & 1.60 & 106.67 & 450.63 & 1.18 & 275.71 & 84.10 & 1.58 & 343.97 & 61.36 & 1.46 \\
\bottomrule
\end{tabular}
\caption{Same as table~\ref{tb:nuenue} for the $b\overline{b}$  channel.}
\label{tb:bbar}
\end{table*}
\end{center}

\begin{center}
\begin{table*}
\begin{tabular}{|r| c c c | c c c |  c c c |  c c c | }
\toprule
 & \multicolumn{6}{c|}{{\bf Annihilation}} & \multicolumn{6}{c|}{{\bf Decay}} \\
 & \multicolumn{3}{c|}{{\bf NFW }} &  \multicolumn{3}{c|}{{\bf Burkert}} & \multicolumn{3}{c|}{{\bf NFW}} &  \multicolumn{3}{c|}{\bf{Burkert}} \\
{$m_\chi$} &  \multirow{2}{*}{$\hat{n}_s$} & {$\langle \sigma v \rangle^{90\%}_{u.l.} $} & \multirow{2}{*}{$z$-score} &  \multirow{2}{*}{$\hat{n}_s$} & {$\langle \sigma v \rangle^{90\%}_{u.l.}$}   & \multirow{2}{*}{$z$-score} &  \multirow{2}{*}{$\hat{n}_s$} & {$\tau^{90\%}_{u.l.}$} & \multirow{2}{*}{$z$-score} &  \multirow{2}{*}{$\hat{n}_s$} & {$\tau^{90\%}_{u.l.}$} & \multirow{2}{*}{$z$-score} \\ 
${\rm [GeV]}$ &   & {$10^{-24} {\rm \;[cm^2]}$} &   &   & {$10^{-24} {\rm \;[cm^2]}$} &  &  &  {$10^{24} {\rm \;[s]}$} &   &    & {$10^{24} {\rm \;[s]}$} & \\
\midrule
100 & 0.01 & 8.93 & 0.00 & 0.00 & 25.72 & 0.00 & - & - & - & - & - & - \\
158 & 108.67 & 14.42 & 0.68 & 0.00 & 20.62 & 0.00 & - & - & - & - & - & - \\
251 & 103.96 & 11.68 & 1.16 & 162.95 & 44.30 & 0.89 & 0.00 & 11.50 & 0.00 & 0.00 & 9.42 & 0.00 \\
398 & 96.87 & 12.03 & 1.45 & 244.93 & 59.29 & 1.69 & 60.08 & 10.59 & 0.18 & 0.01 & 9.22 & 0.00 \\
631 & 78.34 & 13.98 & 1.36 & 267.65 & 78.94 & 2.14 & 338.60 & 10.81 & 1.38 & 446.23 & 7.64 & 1.35 \\
1000 & 156.43 & 12.94 & 2.15 & 325.55 & 55.74 & 2.19 & 633.39 & 14.94 & 2.65 & 783.64 & 11.13 & 2.48 \\
1585 & 105.55 & 11.06 & 1.71 & 210.19 & 46.99 & 1.68 & 555.29 & 24.61 & 2.38 & 730.95 & 17.63 & 2.37 \\
2512 & 100.09 & 12.94 & 1.59 & 99.67 & 38.16 & 0.93 & 403.61 & 44.08 & 1.94 & 512.46 & 32.23 & 1.88 \\
3981 & 37.32 & 9.89 & 0.83 & 83.27 & 44.92 & 0.88 & 237.28 & 82.50 & 1.34 & 328.46 & 57.55 & 1.41 \\
6310 & 41.74 & 12.30 & 1.28 & 66.79 & 49.85 & 0.91 & 128.47 & 140.78 & 0.84 & 210.19 & 90.99 & 1.02 \\
10000 & 39.78 & 16.29 & 1.40 & 57.08 & 62.30 & 0.91 & 133.24 & 176.62 & 1.07 & 171.72 & 124.80 & 0.99 \\
15850 & 38.02 & 22.61 & 1.45 & 54.08 & 89.70 & 0.91 & 104.07 & 246.60 & 1.03 & 117.88 & 182.60 & 0.84 \\
25120 & 37.98 & 35.68 & 1.47 & 51.32 & 133.40 & 0.89 & 92.33 & 289.76 & 1.01 & 97.55 & 220.01 & 0.76 \\
39810 & 36.53 & 58.84 & 1.51 & 48.12 & 199.24 & 0.86 & 87.95 & 313.58 & 1.01 & 96.77 & 234.15 & 0.79 \\
\bottomrule
\end{tabular}
\caption{Same as table~\ref{tb:nuenue} for the $W^+W^-$ channel.}
\label{tb:WW}
\end{table*}
\end{center}

\begin{center}
\begin{table*}
\begin{tabular}{|r| c c c | c c c |  c c c |  c c c | }
\toprule
 & \multicolumn{6}{c|}{{\bf Annihilation}} & \multicolumn{6}{c|}{{\bf Decay}} \\
 & \multicolumn{3}{c|}{{\bf NFW }} &  \multicolumn{3}{c|}{{\bf Burkert}} & \multicolumn{3}{c|}{{\bf NFW}} &  \multicolumn{3}{c|}{\bf{Burkert}} \\
{$m_\chi$} &  \multirow{2}{*}{$\hat{n}_s$} & {$\langle \sigma v \rangle^{90\%}_{u.l.} $} & \multirow{2}{*}{$z$-score} &  \multirow{2}{*}{$\hat{n}_s$} & {$\langle \sigma v \rangle^{90\%}_{u.l.}$}   & \multirow{2}{*}{$z$-score} &  \multirow{2}{*}{$\hat{n}_s$} & {$\tau^{90\%}_{u.l.}$} & \multirow{2}{*}{$z$-score} &  \multirow{2}{*}{$\hat{n}_s$} & {$\tau^{90\%}_{u.l.}$} & \multirow{2}{*}{$z$-score} \\ 
${\rm [GeV]}$ &   & {$10^{-24} {\rm \;[cm^2]}$} &   &   & {$10^{-24} {\rm \;[cm^2]}$} &  &  &  {$10^{24} {\rm \;[s]}$} &   &    & {$10^{24} {\rm \;[s]}$} & \\
\midrule
10 & 97.09 & 217.83 & 0.92 & 0.04 & 363.10 & 0.00 & - & - & - & - & - & - \\
16 & 0.05 & 22.32 & 0.00 & 0.00 & 64.80 & 0.00 & 214.63 & 0.02 & 2.19 & 68.44 & 0.01 & 0.25 \\
25 & 0.00 & 6.48 & 0.00 & 0.00 & 28.54 & 0.00 & 0.00 & 0.16 & 0.00 & 0.00 & 0.13 & 0.00 \\
40 & 0.01 & 5.21 & 0.00 & 29.84 & 26.28 & 0.08 & 0.00 & 1.12 & 0.00 & 0.10 & 0.89 & 0.00 \\
63 & 0.03 & 5.72 & 0.00 & 88.01 & 23.31 & 0.24 & 0.38 & 2.20 & 0.00 & 0.00 & 1.96 & 0.00 \\
100 & 11.57 & 4.24 & 0.07 & 0.00 & 12.74 & 0.00 & 112.26 & 3.57 & 0.20 & 39.05 & 3.01 & 0.06 \\
158 & 0.00 & 3.22 & 0.00 & 0.00 & 10.06 & 0.00 & 1.09 & 8.36 & 0.00 & 0.08 & 7.25 & 0.00 \\
251 & 25.75 & 3.74 & 0.26 & 87.68 & 19.21 & 0.40 & 0.00 & 21.10 & 0.00 & 0.00 & 18.02 & 0.00 \\
398 & 80.11 & 5.40 & 1.14 & 230.48 & 28.64 & 1.41 & 0.00 & 26.18 & 0.00 & 0.00 & 23.62 & 0.00 \\
631 & 69.87 & 6.03 & 1.25 & 267.84 & 37.93 & 2.06 & 281.87 & 22.88 & 0.99 & 315.87 & 17.34 & 0.83 \\
1000 & 162.30 & 5.95 & 2.45 & 384.23 & 30.72 & 2.43 & 631.29 & 28.75 & 2.47 & 787.90 & 21.27 & 2.33 \\
1585 & 129.09 & 5.85 & 2.03 & 272.62 & 26.52 & 1.99 & 647.18 & 44.33 & 2.62 & 814.07 & 32.66 & 2.50 \\
2512 & 29.90 & 3.19 & 0.57 & 144.80 & 21.09 & 1.28 & 512.25 & 77.18 & 2.26 & 649.87 & 56.54 & 2.18 \\
3981 & 23.89 & 3.44 & 0.60 & 82.87 & 19.87 & 0.88 & 319.50 & 148.42 & 1.69 & 415.80 & 106.52 & 1.66 \\
6310 & 39.10 & 5.06 & 1.28 & 56.45 & 20.33 & 0.81 & 151.08 & 290.62 & 0.96 & 230.33 & 194.68 & 1.11 \\
10000 & 25.43 & 5.58 & 1.17 & 38.35 & 22.42 & 0.73 & 111.20 & 430.45 & 0.89 & 151.98 & 300.26 & 0.91 \\
15850 & 23.29 & 7.95 & 1.25 & 30.63 & 29.69 & 0.70 & 79.20 & 643.16 & 0.88 & 91.74 & 469.49 & 0.73 \\
25120 & 22.11 & 11.85 & 1.38 & 25.83 & 44.15 & 0.66 & 60.25 & 832.27 & 0.85 & 59.54 & 638.93 & 0.60 \\
39810 & 19.49 & 19.24 & 1.28 & 22.70 & 70.53 & 0.66 & 48.56 & 945.58 & 0.80 & 44.47 & 733.32 & 0.52 \\
\\
\bottomrule
\end{tabular}
\caption{Same as table~\ref{tb:nuenue} for the $\tau^+\tau^-$channel.}
\label{tb:tautau}
\end{table*}
\end{center}

\begin{center}
\begin{table*}
\begin{tabular}{|r| c c c | c c c |  c c c |  c c c | }
\toprule
 & \multicolumn{6}{c|}{{\bf Annihilation}} & \multicolumn{6}{c|}{{\bf Decay}} \\
 & \multicolumn{3}{c|}{{\bf NFW }} &  \multicolumn{3}{c|}{{\bf Burkert}} & \multicolumn{3}{c|}{{\bf NFW}} &  \multicolumn{3}{c|}{\bf{Burkert}} \\
{$m_\chi$} &  \multirow{2}{*}{$\hat{n}_s$} & {$\langle \sigma v \rangle^{90\%}_{u.l.} $} & \multirow{2}{*}{$z$-score} &  \multirow{2}{*}{$\hat{n}_s$} & {$\langle \sigma v \rangle^{90\%}_{u.l.}$}   & \multirow{2}{*}{$z$-score} &  \multirow{2}{*}{$\hat{n}_s$} & {$\tau^{90\%}_{u.l.}$} & \multirow{2}{*}{$z$-score} &  \multirow{2}{*}{$\hat{n}_s$} & {$\tau^{90\%}_{u.l.}$} & \multirow{2}{*}{$z$-score} \\ 
${\rm [GeV]}$ &   & {$10^{-24} {\rm \;[cm^2]}$} &   &   & {$10^{-24} {\rm \;[cm^2]}$} &  &  &  {$10^{24} {\rm \;[s]}$} &   &    & {$10^{24} {\rm \;[s]}$} & \\
\midrule
10 & 0.01 & 37.44 & 0.00 & 22.31 & 311.43 & 0.13 & - & - & - & - & - & - \\
16 & 0.01 & 8.71 & 0.00 & 0.20 & 47.20 & 0.00 & 186.02 & 0.03 & 2.00 & 184.94 & 0.01 & 0.66 \\
25 & 50.04 & 7.65 & 0.43 & 0.18 & 22.97 & 0.00 & 0.00 & 0.22 & 0.00 & 0.00 & 0.16 & 0.00 \\
40 & 121.75 & 7.04 & 0.80 & 53.54 & 21.68 & 0.15 & 0.16 & 1.26 & 0.00 & 0.00 & 1.10 & 0.00 \\
63 & 15.90 & 3.86 & 0.10 & 91.12 & 18.53 & 0.25 & 0.01 & 2.66 & 0.00 & 0.00 & 2.41 & 0.00 \\
100 & 0.00 & 3.06 & 0.00 & 0.00 & 9.95 & 0.00 & 127.16 & 4.50 & 0.23 & 36.28 & 3.82 & 0.05 \\
158 & 17.97 & 4.08 & 0.12 & 0.00 & 7.25 & 0.00 & 14.22 & 9.80 & 0.03 & 0.00 & 9.91 & 0.00 \\
251 & 34.25 & 3.80 & 0.37 & 49.61 & 14.04 & 0.25 & 0.00 & 28.47 & 0.00 & 0.00 & 22.16 & 0.00 \\
398 & 84.91 & 5.33 & 1.26 & 203.61 & 22.85 & 1.42 & 0.00 & 38.65 & 0.00 & 0.00 & 29.17 & 0.00 \\
631 & 73.61 & 5.88 & 1.47 & 236.07 & 30.91 & 2.10 & 252.54 & 28.76 & 1.00 & 293.35 & 21.39 & 0.86 \\
1000 & 142.61 & 5.17 & 2.06 & 356.56 & 24.50 & 2.42 & 579.06 & 37.10 & 2.41 & 750.01 & 26.76 & 2.35 \\
1585 & 116.00 & 5.18 & 1.82 & 236.63 & 20.55 & 1.88 & 598.32 & 56.02 & 2.60 & 747.97 & 41.54 & 2.46 \\
2512 & 51.84 & 4.16 & 0.95 & 109.62 & 15.56 & 1.08 & 457.70 & 98.58 & 2.19 & 572.94 & 72.77 & 2.08 \\
3981 & 27.95 & 3.52 & 0.77 & 50.88 & 14.11 & 0.61 & 259.20 & 197.61 & 1.53 & 333.22 & 142.71 & 1.48 \\
6310 & 38.13 & 5.25 & 1.44 & 39.28 & 15.53 & 0.66 & 107.25 & 396.86 & 0.78 & 172.73 & 260.16 & 0.94 \\
10000 & 27.37 & 5.90 & 1.45 & 28.31 & 17.74 & 0.64 & 77.02 & 573.12 & 0.71 & 106.84 & 393.47 & 0.72 \\
15850 & 23.98 & 8.59 & 1.44 & 23.38 & 25.14 & 0.64 & 56.53 & 835.23 & 0.75 & 62.00 & 618.02 & 0.59 \\
25120 & 17.36 & 11.99 & 1.21 & 20.73 & 40.82 & 0.63 & 45.37 & 1027.81 & 0.77 & 42.72 & 784.36 & 0.52 \\
39810 & 18.33 & 23.40 & 1.28 & 19.11 & 71.71 & 0.62 & 39.57 & 1055.78 & 0.77 & 33.58 & 787.77 & 0.43 \\
\bottomrule
\end{tabular}
\caption{Same as table~\ref{tb:nuenue} for the $\mu^+\mu^-$channel.}
\label{tb:mumu}
\end{table*}
\end{center}

%\newpage

\bibliography{references}

\end{document}